\documentclass[a4paper,11pt]{article}
\usepackage[utf8]{inputenc}
\usepackage[T1]{fontenc}
\usepackage[english]{babel}
\usepackage[hmargin={32mm,32mm},vmargin={32mm,35mm}]{geometry}
\usepackage{amsmath}
\usepackage{amsfonts}
\usepackage{amssymb}
\usepackage[mathscr]{euscript}
\usepackage{enumerate}
\usepackage{graphicx}
\usepackage{bbm}
\usepackage{color}
\usepackage{xspace}
\usepackage{url}
\usepackage[hidelinks]{hyperref}
\usepackage{caption}
\usepackage{enumitem}
\usepackage{bookmark}
\usepackage{csquotes}
\usepackage[
	bibstyle=phys,
	biblabel=brackets,
	citestyle=numeric-comp,
	sorting=none,
	doi=false,
	eprint=true,
	pageranges=false,
	backend=biber
]{biblatex}

\DeclareFieldFormat[article,inproceedings,inbook]{title}{\mkbibitalic{#1\isdot}}
\AtEveryBibitem{
	\ifentrytype{book}{
		\clearfield{series}
		\clearfield{volume}
	}{}
	\ifentrytype{inbook}{
		\clearfield{pages}
	}{}
}

\newcommand{\bra}[1]{\langle #1 |}
\newcommand{\ket}[1]{|#1\rangle}
\newcommand{\braket}[2]{\langle #1 | #2 \rangle}

\newcommand{\bket}[1]{\bigl|#1\bigr\rangle}

\newcommand{\eg}{e.g.\@\xspace}
\newcommand{\ie}{i.e.\@\xspace}
\newcommand{\Eq}{Eq.\@\xspace}
\newcommand{\Eqs}{Eqs.\@\xspace}
\newcommand{\Fig}{Fig.\@\xspace}

\newcommand{\downup}[2]{_{#1}^{\phantom{#1}#2}}

\newcommand{\R}{\mathbbm{R}}

\newcommand{\Id}{\mathbbm{1}}
\newcommand{\const}{\mathrm{const.}}
\newcommand{\Tr}{\operatorname{Tr}}
\newcommand{\D}{{\cal D}}
\newcommand{\DD}{{\mathscr{D}}}
\newcommand{\V}{{\cal V}}
\newcommand{\Et}{\widetilde E}
\newcommand{\Rt}{\,{}^{(3)}\!R}

\numberwithin{equation}{section}

\begin{document}

\begin{center}

\Large
\textbf{Master constraint approach to quantum-reduced\\loop gravity}

\vspace{16pt}

\large
Ilkka Mäkinen

\normalsize

\vspace{12pt}

National Centre for Nuclear Research \\
Pasteura 7, 02-093 Warsaw, Poland

\vspace{8pt}

ilkka.makinen@ncbj.gov.pl

\end{center}

\renewcommand{\abstractname}{\vspace{-\baselineskip}}

\begin{abstract}
	\noindent We introduce a master constraint operator on the kinematical Hilbert space of loop quantum gravity representing a set of gauge conditions which classically fix the densitized triad to be diagonal. We argue that the master constraint approach provides a natural and systematic way of carrying out the quantum gauge-fixing procedure which underlies the model known as quantum-reduced loop gravity. The Hilbert space of quantum-reduced loop gravity is obtained as a particular space of solutions of the gauge-fixing master constraint operator. We give a concise summary of the fundamental structure of the quantum-reduced framework, and consider several possible extensions thereof, for which the master constraint formulation provides a convenient starting point. In particular, we propose a generalization of the standard Hilbert space of quantum-reduced loop gravity, which may be relevant in the application of the quantum-reduced model to physical situations in which the Ashtekar connection is not diagonal.
\end{abstract}

\tableofcontents

\thispagestyle{empty}

\newpage
\setcounter{page}{1}

\section{Introduction}

Quantum-reduced loop gravity is a physically motivated model of loop quantum gravity (LQG), which is designed to address the crucial question of extracting statements about concrete physical phenomena from the formalism of full loop quantum gravity. From the practical standpoint, a key advantage of the quantum-reduced model is its simplified kinematical structure, which allows for physical applications of the model to be developed at the level of explicit calculations. Quantum-reduced loop gravity was introduced by Alesci and Cianfrani in \cite{Alesci:2012md, Alesci:2013xd} and studied further in several subsequent works, \eg \cite{Alesci:2014uha, Alesci:2014rra, Alesci:2016rmn, Alesci:2016gub, Alesci:2018loi}. The model has found applications in cosmology \cite{Alesci:2017kzc, Alesci:2018qtm, Olmedo:2018ohq, Alesci:2019sni} and in the physics of black holes \cite{Alesci:2019pbs, Alesci:2020zfi, Gan:2022mle}, and its relation with full loop quantum gravity has been studied particularly in \cite{Alesci:2013xya, Makinen:2020rda}.

On a technical level, quantum-reduced loop gravity is based on the idea of implementing in the quantum theory a gauge fixing to a diagonal densitized triad. The gauge conditions select a sector of the kinematical Hilbert space of full loop quantum gravity, which serves as the Hilbert space of the quantum-reduced model. The kinematical structure of quantum-reduced loop gravity is then completed by the definition of the model's basic operators, which can be seen as arising from the action of the corresponding operators of full loop quantum gravity on the aforementioned Hilbert space \cite{Makinen:2020rda}. A distinctive feature of the resulting framework is the considerable simplicity of its quantum states and operators, when compared with their counterparts in the full theory.\footnote{
	As a specific example we may mention the volume operator, which is important in loop quantum gravity not only as a fundamental geometrical observable, but also as a key element in the formulation of the dynamics (the volume operator is involved in any standard definition of the Hamiltonian constraint operator in loop quantum gravity). In full loop quantum gravity, the matrix elements of the volume operator between basis states of the kinematical Hilbert space cannot be computed in explicit form in general, but only in certain simple enough special cases (see \eg \cite{Brunnemann:2007ca, Brunnemann:2007as}). In contrast, the volume operator in quantum-reduced loop gravity acts diagonally on the natural basis states of the Hilbert space of the model.
}

Thus, the choice of gauge which constitutes the fundamental premise of quantum-reduced loop gravity is motivated ultimately by practical considerations: Starting from the framework of full loop quantum gravity, the program initiated by Alesci and Cianfrani is able to establish a model which, on one hand, enjoys a sufficiently simplified kinematical structure to enable concrete calculations about physics to be performed within the model, while on the other hand retaining a clear and definite connection with the formalism of the full theory.

In the original work of Alesci and Cianfrani \cite{Alesci:2012md, Alesci:2013xd}, the gauge conditions for a diagonal triad were enforced in the quantum theory through an intricate procedure involving $U(1)$ holonomies and the formalism of projected spin networks \cite{Dupuis:2010jn}. In the present article we propose the master constraint method, originally introduced to loop quantum gravity by Thiemann \cite{Thiemann:2003zv, Thiemann:2005zg, Giesel:2006uj} in connection with the Hamiltonian constraint, as a natural and straightforward approach for implementing the quantum gauge-fixing conditions underlying quantum-reduced loop gravity. The classical starting point for the construction of a master constraint operator is to assemble the gauge conditions fixing the densitized triad $E^a_i$ to be diagonal, \ie $E^a_i = 0$ for $a\neq i$, into a single master constraint functional $M$, such that the single condition $M=0$ is equivalent to the gauge conditions being satisfied everywhere on the spatial manifold $\Sigma$. Upon quantization, the master constraint is promoted into a constraint operator, which represents the gauge conditions for a diagonal triad on the kinematical Hilbert space of loop quantum gravity.

In the context of quantum-reduced loop gravity and quantum gauge-fixing, an advantage of the master constraint approach is that it results in the gauge conditions being formulated entirely in terms of a well-defined constraint operator. This allows one to largely avoid a certain level of confusion, which is present in much of the early literature of quantum-reduced loop gravity regarding various technical aspects of the framework. In a sense, the present article can therefore be seen as a complement to the earlier article \cite{Makinen:2020rda}, where the status of the fundamental operators of quantum-reduced loop gravity was clarified by establishing a clear relation between these operators and the corresponding operators of full loop quantum gravity. The work presented in this article arguably accomplishes an analogous clarification for the quantum states which form the Hilbert space of quantum-reduced loop gravity.

The material in this article is organized as follows. In section \ref{sec:LQG} we give a brief summary of the basic kinematical framework of loop quantum gravity. This serves to establish our notation and conventions, and to introduce the necessary tools which will be used in the main part of the article. (For a more comprehensive introduction to loop quantum gravity, we refer the reader to any of the standard references, \eg \cite{Ashtekar:2004eh, Rovelli:2004tv, Han:2005km, Thiemann:2007pyv, Rovelli:2014ssa, Ashtekar:2017yom}.) In section \ref{sec:master} we define a master constraint operator representing the gauge conditions which fix the triad to be diagonal on the kinematical Hilbert space of loop quantum gravity. We show that the states spanning the Hilbert space of quantum-reduced loop gravity can be recovered as solutions of our master constraint operator on cubical graphs, with the notion of solutions being understood in the generalized sense\footnote{
	Namely, since the master constraint operator represents a set of constraints which are non-commuting at the level of the quantum theory (although they are compatible with each other classically), one acknowledges that looking for solutions of the constraint equation as an exact equality would be too restrictive of a condition. Instead, (generalized) solutions are searched in the form of states which carry large spin quantum numbers, and which approximately satisfy the constraint equation in the regime of large spins.
}
utilized by Alesci and Cianfrani when originally introducing the quantum-reduced model. 

Section \ref{sec:QRLG} presents a concise but methodical overview of the fundamental structure of quantum-reduced loop gravity, recapitulating the results obtained in section \ref{sec:master} regarding the Hilbert space of the quantum-reduced model, and in the earlier article \cite{Makinen:2020rda} regarding the basic operators of the model. When the Hilbert space of quantum-reduced loop gravity is viewed as a subspace of the kinematical Hilbert space of full loop quantum gravity, operators of the quantum-reduced model arise in a natural way from the action of operators of full loop quantum gravity on the reduced Hilbert space. A given operator of the full theory typically gives rise to a corresponding ''reduced'' operator, which is obtained by discarding certain small terms generated by the action of the full operator on states in the reduced Hilbert space. The reduced operator is a well-defined operator in the framework of the quantum-reduced model, the Hilbert space of the model being preserved by the action of this operator, while from the point of view of full loop quantum gravity, the reduced operator provides a good approximation of the action of the corresponding full theory operator on the reduced Hilbert space.

In section \ref{sec:generalizations} we consider various possible generalizations of the standard formulation of quantum-reduced loop gravity. We discuss the question of $SU(2)$ gauge invariance in the quantum-reduced model, and present two ways in which this question could in principle be addressed. We also introduce an extension of the standard Hilbert space of quantum-reduced loop gravity, which is obtained by splitting each edge of a standard reduced basis state into two ''half-edges'', which can carry arbitrary magnetic quantum numbers at their free ends. We argue that a generalization of the reduced Hilbert space (which may be the generalization proposed in section \ref{sec:half-edges}, or another, so far unknown generalization of a different kind) is likely needed in order to capture all the degrees of freedom contained in a general, not necessarily diagonal Ashtekar connection. Finally, the concluding section \ref{sec:conclusions} summarizes the work presented in this article and briefly comments on possible directions for future work.

\section{A brief review of loop quantum gravity}
\label{sec:LQG}

\subsection{The kinematical Hilbert space}
\label{sec:LQG-states}

The kinematical Hilbert space of loop quantum gravity is formed by the so-called cylindrical functions. (More precisely, the kinematical Hilbert space is defined as the completion of the space of cylindrical functions with respect to the scalar product defined in \Eq \eqref{eq:ALproduct} and in the subsequent text.) A function cylindrical with respect to a graph $\Gamma$ is a complex-valued function of the form
\begin{equation}
	\Psi_\Gamma(h_{e_1}, \dots, h_{e_N}).
	\label{eq:Psi-cyl}
\end{equation}
The graph $\Gamma$ consists of $N$ edges $e_1, \dots, e_N$, which are assumed to be oriented and embedded in the spatial manifold $\Sigma$. The arguments of the cylindrical function \eqref{eq:Psi-cyl} are $SU(2)$-valued group elements, one for each edge of the graph.

The group elements $h_e$ are typically referred to as holonomies, due to their origin as holonomies (path-ordered exponentials) of the Ashtekar--Barbero connection in the classical theory. In the quantum theory the holonomies are assumed to satisfy certain algebraic properties with respect to the edge label $e$, which are consistent with the classical interpretation of the holonomy as a parallel propagator. These properties are
\begin{align}
	h_{e^{-1}} &= h_e^{-1}, \label{eq:h^-1} \\
	h_{e_2\circ e_1} &= h_{e_2}h_{e_1}, \label{eq:h2h1} \\
	h_p &= \Id,
	\label{}
\end{align}
where $e^{-1}$ denotes the edge $e$ taken with the opposite orientation, $e_2\circ e_1$ denotes the combined edge obtained by taking $e_1$ followed by $e_2$ (it is assumed that the endpoint of $e_1$ coincides with the beginning point of $e_2$), and $p$ denotes an edge consisting of a single point.

A scalar product on the space of cylindrical functions is given by the Ashtekar--Lewandowski scalar product \cite{Ashtekar:1993wf, Ashtekar:1994mh}. This is defined as follows. For two functions cylindrical with respect to the same graph $\Gamma$, we set
\begin{equation}
	\braket{\Psi_\Gamma}{\Psi'_\Gamma} = \int dh_1\cdots dh_N\,\overline{\Psi_\Gamma(h_1,\dots,h_N)}\Psi'_\Gamma(h_1,\dots,h_N)
	\label{eq:ALproduct}
\end{equation}
where $dh$ denotes the normalized Haar measure of $SU(2)$. To extend the definition to two functions cylindrical with respect to two different graphs $\Gamma_1$ and $\Gamma_2$, consider any graph $\Gamma_{12}$ that contains both $\Gamma_1$ and $\Gamma_2$ as subgraphs. By introducing a trivial dependence on the group elements associated with the additional edges, each of the two functions in question can be seen as a cylindrical function on $\Gamma_{12}$ (see \eg \cite{Ashtekar:2004eh}), and their scalar product is then defined by applying \Eq \eqref{eq:ALproduct} on $\Gamma_{12}$. (The normalization of the Haar measure implies that the result is independent of the choice of the graph $\Gamma_{12}$.)

It follows from the Peter--Weyl theorem that a basis on the space of cylindrical functions can be constructed from the irreducible representation matrices of $SU(2)$ (Wigner matrices) $D^{(j)}_{mn}(h)$. An orthonormal basis on the space of functions cylindrical with respect to a graph $\Gamma$ is given by the functions
\begin{equation}
	\prod_{e\in\Gamma} \DD^{(j_e)}_{m_en_e}(h_e)
	\label{eq:basis_kin}
\end{equation}
where the quantum numbers $\{j_e, m_e, n_e\}$ range over all their possible values, and the notation
\begin{equation}
	\DD^{(j)}_{mn}(h_e) = \sqrt{2j+1}D^{(j)}_{mn}(h_e)
	\label{}
\end{equation}
has been introduced for the normalized\footnote{
	In the sense of the Haar measure, \ie $\int dh\,\bigl|\DD^{(j)}_{mn}(h)\bigr|^2 = 1$.
}
matrix elements of the Wigner matrices.

An important subspace of the space of all cylindrical functions is the space of functions invariant under local $SU(2)$ gauge transformations. (Classically such transformations correspond to internal rotations of the densitized triad $E^a_i$.) Under a gauge transformation given by a gauge function $g(x)\in SU(2)$, the holonomy transforms as
\begin{equation}
	h_e \to g\bigl(t(e)\bigr)h_eg^{-1}\bigl(s(e)\bigr)
	\label{eq:g*h}
\end{equation}
where $s(e)$ and $t(e)$ stand for the beginning point (source) and endpoint (target) of the edge $e$. The space of gauge invariant cylindrical functions on a graph $\Gamma$ is spanned by cylindrical functions of the form  
\begin{equation}
	\biggl(\prod_{v\in\Gamma} \iota_v\biggr)\cdot\biggl(\prod_{e\in\Gamma} \DD^{(j_e)}(h_e)\biggr).
	\label{eq:spinnetwork}
\end{equation}
These are the well-known spin network functions of loop quantum gravity \cite{Rovelli:1995ac, Baez:1995md}. In the function \eqref{eq:spinnetwork}, an invariant tensor of $SU(2)$ (intertwiner) $\iota_v$ of an appropriate index structure is associated with each node $v$ of the graph, and the dot indicates a complete contraction of magnetic indices according to the pattern dictated by the graph. More precisely, suppose a node $v$ contains $M$ edges oriented inwards and carrying spins $j_1, \dots, j_M$, and $N$ edges oriented outwards and carrying spins $j_1', \dots, j_N'$. Then the tensors $\iota_v$ assigned to the node form a basis of the intertwiner space
\begin{equation}
	{\rm Inv}_{SU(2)}\Bigl({\cal H}_{j_1}\otimes\cdots\otimes{\cal H}_{j_M}\otimes{\cal H}_{j_1'}^*\otimes\cdots\otimes{\cal H}_{j_N'}^*\Bigr),
	\label{eq:intertwinerspace}
\end{equation}
where ${\cal H}_j^*$ is the dual space, and ${\rm Inv}_{SU(2)}$ denotes the $SU(2)$-invariant subspace. The condition for $\iota_v$ to be invariant under $SU(2)$ is to be understood in an appropriate generalized sense, which takes into account the index structure of the intertwiner. Being an element of the space \eqref{eq:intertwinerspace}, $\iota_v$ is a tensor with $M$ lower indices (in the representations $j_1, \dots, j_M$) and $N$ upper indices (in the representations $j_1', \dots, j_N'$), and is required to satisfy
\begin{equation}
	D^{(j_1')}(g^{-1})\cdots D^{(j_N')}(g^{-1})\iota_vD^{(j_1)}(g)\cdots D^{(j_M)}(g) = \iota_v
	\label{}
\end{equation}
where $g$ is an arbitrary element of $SU(2)$, and the action of the representation matrices on $\iota_v$ is given by contraction of magnetic indices.

\subsection{Kinematical operators}
\label{sec:LQG-operators}

The elementary operators of loop quantum gravity are the holonomy and flux operators. These are the quantum operators corresponding to holonomies of the Ashtekar--Barbero connection along one-dimensional curves, and fluxes of the densitized triad through two-dimensional surfaces.

The holonomy operator associated to an edge $e$ acts on cylindrical functions by multiplication:
\begin{equation}
	\widehat{D^{(j)}_{mn}(h_e)}\Psi_\Gamma(h_{e_1}, \dots, h_{e_N}) = D^{(j)}_{mn}(h_e)\Psi_\Gamma(h_{e_1}, \dots, h_{e_N}).
	\label{eq:h*Psi}
\end{equation}
The form of the resulting state depends on whether the edge $e$ is contained among the edges $e_1, \dots, e_N$ of the graph $\Gamma$. For the work carried out in the present article, the relevant case is that $e$ coincides with one of the edges $e_1, \dots, e_N$. In this case, the function on the right-hand side of \Eq \eqref{eq:h*Psi} is still a cylindrical function on the same graph $\Gamma$. The explicit form of this function in the basis \eqref{eq:basis_kin} can be found by using the $SU(2)$ Clebsch--Gordan series
\begin{equation}
	D^{(j_1)}_{m_1n_1}(h_e) D^{(j_2)}_{m_2n_2}(h_e) = \sum_j C^{(j_1\;j_2\;j)}_{m_1\;m_2\;m_1+m_2}C^{(j_1\;j_2\;j)}_{n_1\;n_2\;n_1+n_2}D^{(j)}_{m_1+m_2\; n_1+n_2}(h_e)
	\label{eq:CGseries}
\end{equation}
to couple the holonomies on the edge $e$; here $C^{(j_1\;j_2\;j)}_{m_1\;m_2\;m}$ are the Clebsch--Gordan coefficients of $SU(2)$. (In calculations where the phases of the Clebsch--Gordan coefficients are relevant, we assume that they have been fixed according to the Condon--Shortley convention.)

Consider then the flux operator associated to a surface $S$. The action of the flux operator on a holonomy $D^{(j)}(h_e)$ depends on the relative location of the edge $e$ with respect to the surface $S$. Assuming for simplicity that there is a single point of intersection between the surface $S$ and an edge $e$, the action of the flux operator is given by
\begin{equation}
	\hat E_i(S)D^{(j)}(h_e) = i\nu(S, e)\times\begin{cases}
		\dfrac{1}{2}D^{(j)}(h_e)\tau_i^{(j)} & \text{$s(e)$ lies on $S$} \\[1.7ex] 
		\dfrac{1}{2}\tau_i^{(j)} D^{(j)}(h_e) & \text{$t(e)$ lies on $S$} \\[1.7ex]
		D^{(j)}(h_{e_1})\tau_i^{(j)}D^{(j)}(h_{e_2}) & \text{$S$ intersects $e$ at an interior point}
	\end{cases}
	\label{eq:E*psi}
\end{equation}
Here $\nu(S, e)$ denotes the relative orientation of the surface and the edge at the intersection point, and $\nu(S, e) = 0$ if $e$ intersects $S$ tangentially. In the case that $S$ intersects $e$ at an interior point, $e_1$ and $e_2$ denote the two segments into which $e$ is divided by $S$.  Moreover, $\tau^{(j)}_i$ are the anti-Hermitian generators of $SU(2)$ in the spin-$j$ representation. The matrix elements of $\tau^{(j)}_i$ are given in terms of the matrix elements of the angular momentum operator as
\begin{equation}
	(\tau_i^{(j)})_{mn} = -i\bra{jm}J_i\ket{jn}.
	\label{eq:tau-def}
\end{equation}
In the standard basis, which diagonalizes the $z$-component of angular momentum, we have the following explicit expressions:
\begin{align}
	(\tau_x^{(j)})_{mn} &= -iA_+(j, n)\delta_{m, n+1} - iA_-(j, n)\delta_{m, n-1} \label{eq:tau_x} \\
	(\tau_y^{(j)})_{mn} &= -A_+(j, n)\delta_{m, n+1} + A_-(j, n)\delta_{m, n-1} \label{eq:tau_y} \\
	(\tau_z^{(j)})_{mn} &= -im\delta_{mn} \label{eq:tau_z}
\end{align}
where
\begin{equation}
	A_\pm(j, m) = \frac{1}{2}\sqrt{j(j+1) - m(m\pm 1)}.
	\label{eq:A_jm}
\end{equation}
The action of the flux operator is extended on the basis states \eqref{eq:basis_kin} by the Leibniz rule, \ie each edge in the product \eqref{eq:basis_kin} which intersects the surface $S$ contributes a term of the form \eqref{eq:E*psi} to the action of the operator.

The volume operator \cite{Rovelli:1994ge, Ashtekar:1997fb} is a quantization of the classical functional $\int d^3x\,\sqrt{|\det E|}$. When acting on a cylindrical function, the volume operator picks up contributions from the nodes of the graph, and its action has the generic form
\begin{equation}
	\hat V\ket{\Psi_\Gamma} = \sum_{v\in\Gamma} \hat V_v\ket{\Psi_\Gamma}
	\label{eq:V}
\end{equation}
with
\begin{equation}
	\hat V_v = \sqrt{|\hat q_v|}.
	\label{eq:V_v}
\end{equation}
Here $\hat q_v$ is an operator which can be expressed in explicit form in terms of left- and right-invariant vector fields of $SU(2)$. In contrast, no such explicit expression is available for the operator $\hat V_v$, which must in general be defined implicitly through its spectral decomposition. In this article we will work exclusively with the Ashtekar--Lewandowski version of the volume operator \cite{Ashtekar:1997fb}, and will refer to this operator simply as the volume operator. 

In the construction of operators in loop quantum gravity, one often has to deal with negative powers of the volume element $\sqrt{|\det E|}$ appearing in classical functionals which one wishes to quantize. These factors cannot be quantized simply by using the inverse of the volume operator; the volume operator is not invertible, as its spectrum contains the eigenvalue zero. In the literature of loop quantum gravity, this problem is frequently circumvented by using the so-called Tikhonov regularization of the inverse volume operator (examples of the use of this operator can be found \eg in  \cite{Bianchi:2008es, Assanioussi:2020fsz, Varadarajan:2021zrk}). The regularized inverse volume operator can be defined as
\begin{equation}
	\widehat{\V_v^{-1}} = \lim_{\epsilon\to 0} \frac{\hat V_v}{\hat V_v^2 + \epsilon^2}.
	\label{eq:V^-1}
\end{equation}
Note that the operator \eqref{eq:V^-1} shares the same eigenstates as the operator $\hat V_v$, and each eigenvalue of $\widehat{\V_v^{-1}}$ is the inverse of the corresponding eigenvalue of $\hat V_v$, except the zero eigenvalue of $\hat V_v$ remains a zero eigenvalue of $\widehat{\V_v^{-1}}$. Hence, given a complete set of eigenstates of the volume operator, the spectral decomposition of the regularized inverse volume is given by
\begin{equation}
	\widehat{\V_v^{-1}} = \sum_{\lambda\neq 0} \lambda^{-1}\ket{\lambda}\bra{\lambda}
	\label{eq:V^-1_decomposition}
\end{equation}
where $\ket{\lambda}$ denotes an eigenstate with eigenvalue $\lambda$: $\hat V_v\ket{\lambda} = \lambda\ket{\lambda}$.

Another operator which is often used in technical constructions in loop quantum gravity is the so-called parallel transported flux operator (this operator is also known as the gauge covariant flux operator in the loop quantum gravity literature). The parallel transported flux operator is a quantization of the classical functional
\begin{equation}
	\Et_i(S, x_0) = -2\Tr\Bigl(\tau_i\Et(S, x_0)\Bigr)
	\label{eq:Et-def1}
\end{equation}
where
\begin{equation}
	\Et(S, x_0) = \int_S dS_a\,h_{x_0,x}E^a_i(x)\tau^i h^{-1}_{x_0,x}
	\label{eq:Et-def2}
\end{equation}
is a matrix-valued generalized flux variable. In \Eq \eqref{eq:Et-def2}, $dS_a = \tfrac{1}{2}\epsilon_{abc}dx^b\wedge dx^c$ denotes the natural area element on the surface $S$, and $h_{x_0,x} \equiv h_{p(x_0, x)}$ are (classical) holonomies associated to a family of paths $p(x_0, x)$. For each point $x$ on the surface, $p(x_0, x)$ is a path connecting $x$ to a fixed point $x_0$, which may lie on $S$ or outside of it. Apart from their endpoints being fixed, the paths $p(x_0, x)$ can otherwise be chosen in principle arbitrarily, with different choices leading to different, possibly inequivalent implementations of the parallel transported flux operator.

The action of the parallel transported flux operator on a holonomy is discussed \eg in section 3.3 of \cite{Lewandowski:2021iun}. For the purposes of the present article, it is enough to consider the case where the surface $S$ intersects the edge $e$ at a single interior point of $e$. In this case, the action of the operator can be expressed in the form
\begin{equation}
	\hat{\Et_i}(S, x_0)D^{(j)}(h_e) = i\nu(S, e)D^{(j)}(h_{e_2})D^{(j)}\bigl(h_{x_0, x_+}^{-1}\bigr)\tau_i^{(j)}D^{(j)}\bigl(h_{x_0, x_+}\bigr)D^{(j)}(h_{e_1})
	\label{eq:Et-action}
\end{equation}
where $x_+$ denotes the point of intersection between $e$ and $S$, and, as in \Eq \eqref{eq:E*psi}, $e_1$ and $e_2$ are the two segments into which $e$ is divided by the intersection with $S$. We can see that, just as the action of the standard flux operator inserts the generator $\tau^{(j)}_i$ at the point where the surface intersects the edge -- see \Eq \eqref{eq:E*psi} -- so does the action of the parallel transported flux operator insert the matrix
\begin{equation}
	D^{(j)}\bigl(h_{x_0, x_+}^{-1}\bigr)\tau_i^{(j)}D^{(j)}\bigl(h_{x_0, x_+}\bigr),
	\label{}
\end{equation}
which is simply the generator $\tau^{(j)}_i$ ''rotated'' by the adjoint action of the $SU(2)$ group element $h_{x_0, x_+}$.

\section{Master constraint for diagonal gauge}
\label{sec:master}

\subsection{The master constraint approach}

In this section we propose a constraint operator which represents the gauge conditions
\begin{equation}
	E^a_i = 0 \qquad \text{for $i\neq a$}
	\label{eq:E_off_diag=0}
\end{equation}
on the kinematical Hilbert space of loop quantum gravity. (Our convention is that the spatial index takes the values $a = x, y, z$ while the internal index takes the values $i=1, 2, 3$; however, the values $1, 2, 3$ are identified respectively with $x, y, z$ and are often used interchangeably with each other.) In the classical theory, the gauge conditions \eqref{eq:E_off_diag=0} can be realized by fixing (partially\footnote{
		The gauge fixing is partial because there exist some residual diffeomorphisms which preserve the gauge \eqref{eq:E_off_diag=0}. These are the diffeomorphisms which, when interpreted as coordinate transformations on the spatial manifold $\Sigma$, have the form
		\begin{equation*}
			(x, y, z) \to \bigl(X(x), Y(y), Z(z)\bigr)
		\end{equation*}
		where each new coordinate is a function of the corresponding old coordinate only.
}) the gauge freedom corresponding to the Gauss and diffeomorphism constraints. The different gauge conditions classically commute among themselves, $\{E^a_i, E^b_j\} = 0$, which presents no obstruction against imposing the gauge \eqref{eq:E_off_diag=0}. On the other hand, in the quantum theory the different components of the flux operator $\hat E_i(S)$ do not commute with each other in general. Hence the procedure of representing the gauge conditions \eqref{eq:E_off_diag=0} individually as constraint operators and solving them one by one is inherently ambiguous, and one would expect \eg that the eventual space of solutions may depend on the order in which the individual constraints are imposed.

To construct a constraint operator which treats the constraints \eqref{eq:E_off_diag=0} in a symmetric manner, we turn to the master constraint approach, which was introduced to loop quantum gravity by Thiemann \cite{Thiemann:2003zv, Thiemann:2005zg, Giesel:2006uj} as a method for dealing with the quantization of the Hamiltonian constraint. In this approach one begins by defining a master constraint functional $M$ on the classical phase space, such that the condition $M=0$ is equivalent to the gauge conditions \eqref{eq:E_off_diag=0} being satisfied at every point of the spatial manifold $\Sigma$. A master constraint operator $\hat M$ corresponding to the classical functional $M$ is then constructed on the kinematical Hilbert space of loop quantum gravity, and solutions of the constraint are defined as states satisfying the condition
\begin{equation}
	\hat M\ket{\Psi} = 0.
	\label{eq:M*Psi=0}
\end{equation}
In our case, it turns out that the condition \eqref{eq:M*Psi=0} must be interpreted in a certain approximate sense, instead of imposing it as an exact equation. Since the different gauge conditions \eqref{eq:E_off_diag=0} do not commute with each other as quantum operators, one cannot expect to be able to find a large family of states which satisfy the constraint equation \eqref{eq:M*Psi=0} exactly. Following the approach originally taken by Alesci and Cianfrani in their work on quantum-reduced loop gravity \cite{Alesci:2013xd, Alesci:2013xya, Alesci:2016gub}, we understand (generalized) solutions of the condition \eqref{eq:M*Psi=0} to be states which are characterized by large values of the spin quantum numbers, and which approximately satisfy \Eq \eqref{eq:M*Psi=0} in the large spin regime, such that the norm of the state $\hat M\ket{\Psi}$ approaches zero in the formal limit $j\to\infty$.

\subsection{Definition of the master constraint operator}
\label{sec:master_def}

Consider the constraint function
\begin{equation}
	\mu^2 = (E^x_2)^2 + (E^x_3)^2 + (E^y_1)^2 + (E^y_3)^2 + (E^z_1)^2 + (E^z_2)^2.
	\label{eq:mu^2}
\end{equation}
Clearly the condition $\mu^2(x) = 0$ is fulfilled if and only if all the gauge conditions \eqref{eq:E_off_diag=0} are satisfied at the point $x$. A constraint which encodes the gauge conditions at every point on the spatial manifold $\Sigma$ can be obtained by integrating the constraint \eqref{eq:mu^2} over $\Sigma$. We divide $\mu^2$ by $\sqrt q$, with $q = \det(q_{ab})$ being the determinant of the spatial metric, to construct an integrand of density weight $+1$. Integrating the resulting expression over $\Sigma$ then gives rise to the master constraint functional
\begin{equation}
	M = \int_\Sigma d^3x\,\frac{\mu^2}{\sqrt q}.
	\label{eq:M_cl}
\end{equation}
The integrand in \Eq \eqref{eq:M_cl} is a positive definite function. Therefore, in the classical theory, the single constraint equation $M = 0$ is equivalent to the gauge conditions \eqref{eq:E_off_diag=0} being satisfied everywhere on $\Sigma$.

Our task now is to write down an operator corresponding to the classical functional \eqref{eq:M_cl} on the kinematical Hilbert space of loop quantum gravity. To this end, it is useful to add and subtract the squared diagonal components of $E^a_i$ in $\mu^2$, which allows us to rewrite \Eq \eqref{eq:M_cl} as
\begin{equation}
	M = \int_\Sigma d^3x\,\sum_a \frac{\sum_i (E^a_i)^2 - (E^a_a)^2}{\sqrt q}.
	\label{eq:M_rewritten}
\end{equation}
To construct an operator corresponding to the expression \eqref{eq:M_rewritten}, we make use of a system of cubical cells and surfaces analogous to the one employed \eg in the construction of the Ashtekar--Lewandowski volume operator in \cite{Ashtekar:1997fb}. The spatial manifold $\Sigma$ is partitioned into cubical cells, and for concreteness we let each cell have edge length $\epsilon$ and coordinate volume $\epsilon^3$. Inside any given cell $\Box$, we set up an assembly of three surfaces $S^a(\Box)$, $a = x, y, z$. Furthermore, we assume the existence of a fixed Cartesian background coordinate system, and we require that the faces of the cells $\Box$ and the surfaces $S^a(\Box)$ are aligned with the coordinate directions defined by this coordinate system, such that the background coordinate $x^a = \rm const$ on the surface $S^a(\Box)$. When considering the action of the resulting quantum operator on a cylindrical function defined on a graph $\Gamma$, we assume that the partition is adapted to the graph, by which we mean that the following two conditions must be satisfied: (1) The parameter $\epsilon$ is taken small enough to ensure that each cell $\Box$ contains at most a single node of the graph; (2) If a cell $\Box$ contains a node of the graph, the node coincides with the single point of intersection between the surfaces $S^a(\Box)$. For later use, denote this point of intersection by $p_\Box$, \ie $p_\Box = S^x(\Box) \cap S^y(\Box) \cap S^z(\Box)$.

With the above preparations in place, we now turn to defining an operator corresponding to the functional \eqref{eq:M_rewritten}. The form of this operator is postulated as follows. We write the integral over $\Sigma$ as a sum of integrals over the cells $\Box$ and, for each cell $\Box$, quantize every instance of the densitized triad $E^a_i$ as the flux operator $\hat E_i\bigl(S^a(\Box)\bigr)$, and the factor $1/\sqrt q$ as the regularized inverse volume operator $\widehat{\V_{p_\Box}^{-1}}$ of \Eq \eqref{eq:V^-1} associated to the point $p_\Box$. If the cell $\Box$ does not contain a node of the graph of the cylindrical function on which the operator acts, the action of $\widehat{\V_{p_\Box}^{-1}}$ gives zero, due to the fact that the volume operator $\hat V_v$ has a non-vanishing action only on nodes of valence three or higher. For a cell which contains a node, the action of $\widehat{\V_{p_\Box}^{-1}}$ will in general be non-zero, since we have assumed that the point $p_\Box$ coincides with the node inside $\Box$. Hence the action of the master constraint operator on a given cylindrical function will pick up non-trivial contributions only from cells which contain a node, \ie it has the form
\begin{equation}
	\hat M\ket{\Psi_\Gamma} = \sum_{v\in\Gamma} \hat M_v\ket{\Psi_\Gamma}.
	\label{eq:M}
\end{equation}
Since the operator $\widehat{\V_v^{-1}}$ does not commute with the flux operator, a factor ordering has to be specified. We assume that the inverse volume operator is ordered to the left of the flux operators.\footnote{
	The outcome of the derivation performed in section 3.3 is not affected by this choice. The states obtained there as solutions of the master constraint operator are also solutions of the constraint where the operator $\widehat{{\cal V}_v^{-1}}$ is ordered to the right.
}
With this choice, the operator $\hat M_v$ is given by the expression
\begin{equation}
	\hat M_v = \widehat{{\cal V}_v^{-1}} \sum_a \Bigl(\hat A\bigl(S^a(v)\bigr)^2 - \hat E_a\bigl(S^a(v)\bigr)^2\Bigr),
	\label{eq:M_v}
\end{equation}
where we now use the notation $S^a(v)$ to denote the three surfaces placed within the cell containing the node $v$. In \Eq \eqref{eq:M_v} the sum runs over $a = x, y, z$, and
\begin{equation}
	\hat A(S) = \sqrt{\hat E_i(S)\hat E_i(S)}
	\label{}
\end{equation}
is the area operator associated to the surface $S$. This concludes the definition of the master constraint operator corresponding to the classical master constraint \eqref{eq:M_cl}.

\subsection{A family of solutions}
\label{sec:solutions}

\begin{figure}[t]
	\centering
	\includegraphics[scale=0.18]{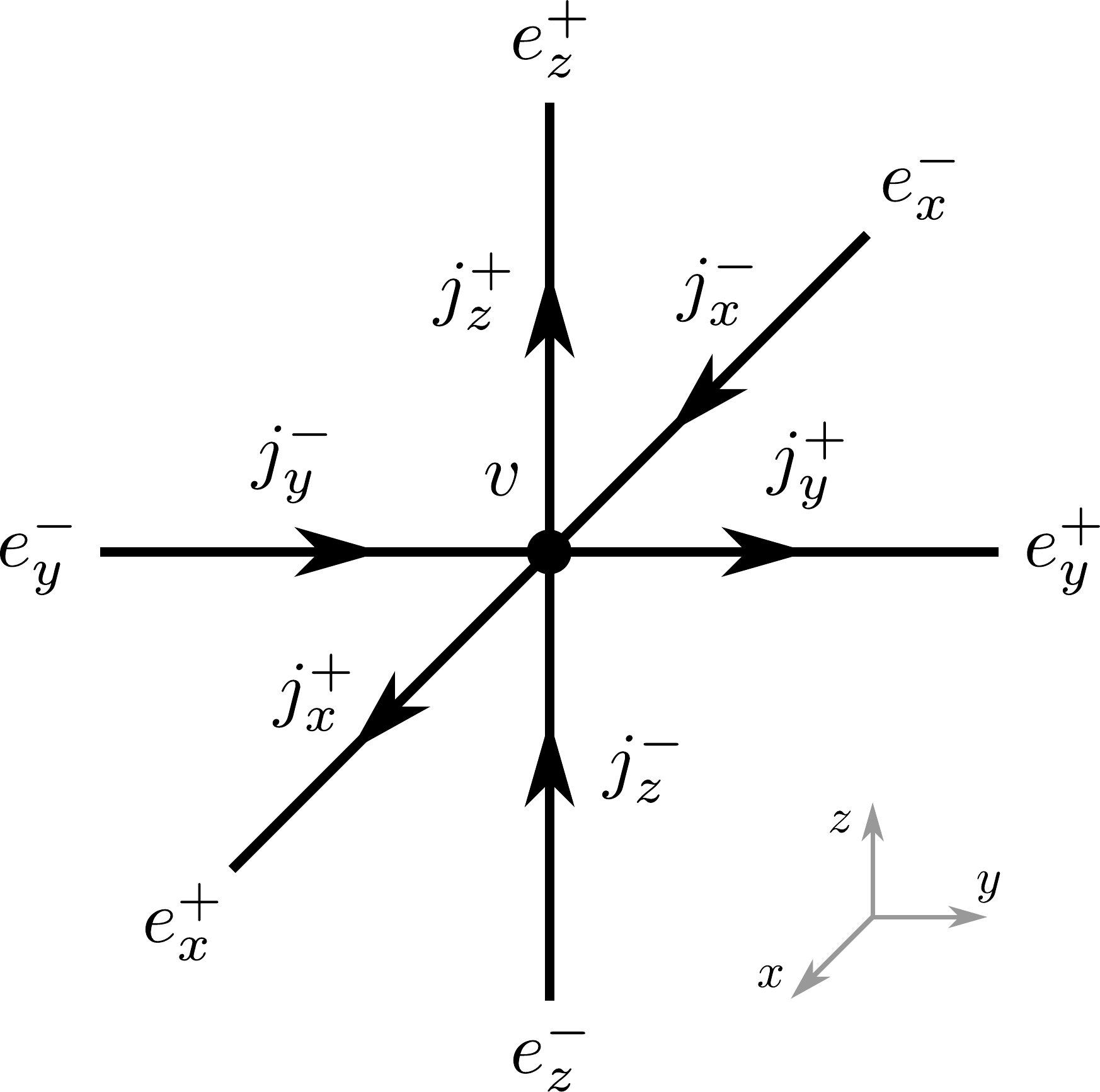}
	\caption{Labeling of the edges and spin quantum numbers for a generic node of a cubical spin network graph. The subscript $i=x, y, z$ in the notation $e_i^\pm$ indicates that the edge is aligned with the $x^a$-coordinate direction of the background coordinate system. The superscript $-$ or $+$ shows whether the edge lies before or after the node $v$ (in the sense defined by the positive direction of the $x^i$-coordinate axis.)}
	\label{fig:node}
\end{figure}

Following the work of Alesci and Cianfrani \cite{Alesci:2013xd, Alesci:2013xya, Alesci:2016gub}, we look for solutions of the master constraint operator defined by \Eqs \eqref{eq:M} and \eqref{eq:M_v} among cylindrical functions based on cubical graphs. By a cubical graph we mean a graph whose nodes are six-valent, and whose edges are aligned with the Cartesian coordinate directions defined by the background coordinate system used in the construction of the master constraint operator in the preceding section. In the calculation which follows, we consider a fixed but arbitrary cubical graph; hence the analysis presented in this section applies to any cubical graph.

Since the operator $\hat M$ acts locally at each node of the graph, we may focus our attention on a single node $v$. We label the edges incident on the node as indicated in \Fig \ref{fig:node}, and we assume that the orientation of each edge $e_i^\pm$ agrees with the positive direction of the corresponding background coordinate axis. Consider now the action of the operator $\hat M_v$ on the generic basis state
\begin{equation}
	\prod_{i=x, y, z} \DD^{(j_i^+)}_{m_i^+n_i^+}\bigl(h_{e_i^+}\bigr)\DD^{(j_i^-)}_{m_i^-n_i^-}\bigl(h_{e_i^-}\bigr)
	\label{eq:node-basis-state}
\end{equation}
associated to the node $v$. For the moment, let us ignore the inverse volume operator in \Eq \eqref{eq:M_v} and focus on the terms in the sum over $a$. Each of these terms will act only on two of the six holonomies in the state \eqref{eq:node-basis-state}, since the edges corresponding to the remaining four holonomies are tangential to the surface $S^a(v)$. For instance, the term with $a=z$ will act only on the factor
\begin{equation}
	\DD^{(j_z^+)}_{m_z^+n_z^+}\bigl(h_{e_z^+}\bigr)\DD^{(j_z^-)}_{m_z^-n_z^-}\bigl(h_{e_z^-}\bigr)
	\label{eq:node-factor-z}
\end{equation}
while leaving the holonomies of the edges in the $x$- and $y$-directions untouched. Using \Eqs \eqref{eq:tau-def}--\eqref{eq:tau_z}, a straightforward calculation shows that the action of this term on the state \eqref{eq:node-factor-z} gives
\begin{align}
	&\Bigl(\hat A\bigl(S^z(v)\bigr)^2 - \hat E_z\bigl(S^z(v)\bigr)^2\Bigr) \DD^{(j_z^+)}_{m_z^+n_z^+}\bigl(h_{e_z^+}\bigr)\DD^{(j_z^-)}_{m_z^-n_z^-}\bigl(h_{e_z^-}\bigr) \notag \\
	&= \frac{1}{4}\Bigl(j_z^+(j_z^++1) + j_z^-(j_z^-+1) - (n_z^+)^2 - (m_z^-)^2\Bigr) \DD^{(j_z^+)}_{m_z^+n_z^+}\bigl(h_{e_z^+}\bigr)\DD^{(j_z^-)}_{m_z^-n_z^-}\bigl(h_{e_z^-}\bigr) \notag \\
	&\qquad + A_+\bigl(j_z^+, n_z^+\bigr)A_-\bigl(j_z^-, m_z^-\bigr) \DD^{(j_z^+)}_{m_z^+,n_z^+ +1}\bigl(h_{e_z^+}\bigr)\DD^{(j_z^-)}_{m_z^- -1, n_z^-}\bigl(h_{e_z^-}\bigr) \notag \\
	&\qquad + A_-\bigl(j_z^+, n_z^+\bigr)A_+\bigl(j_z^-, n_z^-\bigr)\DD^{(j_z^+)}_{m_z^+, n_z^+ -1}\bigl(h_{e_z^+}\bigr)\DD^{(j_z^-)}_{m_z^- +1, n_z^-}\bigl(h_{e_z^-}\bigr).
	\label{eq:M_v-action-z}
\end{align}
Note that, since the magnetic quantum numbers in a given representation $j$ are constrained by $-j \leq m, n \leq j$, the coefficient of the diagonal term on the right-hand side is strictly positive, so the right-hand side as a whole cannot be equal to zero for any values of $m_z^-$, $n_z^+$. This reflects the already anticipated fact that we should generally not expect to be able to find a large class of states satisfying the constraint equation $\hat M\ket{\Psi} = 0$ exactly; the best we can do is to look for states which solve the constraint in an appropriate approximate sense.

Consider the choice
\begin{align}
	n_z^+ &= j_z^+, \label{eq:n=j} \\
	m_z^- &= j_z^-. \label{eq:m=j}
\end{align}
This choice of the magnetic quantum numbers eliminates the off-diagonal terms on the right-hand side of \Eq \eqref{eq:M_v-action-z}, and cancels the terms quadratic in $j$ in the coefficient of the diagonal term. Hence we are left with
\begin{align}
	\Bigl(\hat A\bigl(S^z(v)\bigr)^2 - \hat E_z\bigl(S^z(v)\bigr)^2\Bigr)&\DD^{(j_z^+)}_{m_z^+j_z^+}\bigl(h_{e_z^+}\bigr)\DD^{(j_z^-)}_{j_z^-n_z^-}\bigl(h_{e_z^-}\bigr) \notag \\
	= \frac{1}{4}\bigl(j_z^+ + j_z^-\bigr)&\DD^{(j_z^+)}_{m_z^+j_z^+}\bigl(h_{e_z^+}\bigr)\DD^{(j_z^-)}_{j_z^-n_z^-}\bigl(h_{e_z^-}\bigr).
	\label{}
\end{align}
(In fact, for fixed values of the spin quantum numbers $j_z^\pm$, the choice \eqref{eq:n=j}--\eqref{eq:m=j} leads to the smallest possible value of the norm of the state on the right-hand side of \Eq \eqref{eq:M_v-action-z}.)

Let us at this point address the terms with $a=x$ and $a=y$ in \Eq \eqref{eq:M_v}, which will act on the holonomies corresponding to $i=x$ and $i=y$ in the state \eqref{eq:node-basis-state}. The above calculations were carried out in the standard basis, in which the generator $\tau^{(j)}_z$ is diagonal. These calculations will go through without modification for the $x$- and $y$-directions as well, provided that we do not use the standard basis, but an appropriately constructed basis which diagonalizes the corresponding generator $\tau^{(j)}_x$ or $\tau^{(j)}_y$. This basis is given by
\begin{equation}
	\ket{jm}_i = D^{(j)}(g_i)\ket{jm}
	\label{eq:jm_i}
\end{equation}
where $i = x$ or $y$, and $g_i$ denotes an $SU(2)$ element describing a rotation which rotates the $z$-axis into the $i$-axis. This rotation is of course not uniquely determined by the condition that the $z$-axis is rotated into the $i$-axis. We fix the ambiguity by requiring that the rotation corresponds to a cyclic permutation of the coordinate axes, so that the triple $(x, y, z)$ is transformed into $(y, z, x)$ or $(z, x, y)$. (This choice ensures that \Eqs \eqref{eq:tau_x}--\eqref{eq:tau_z} are valid in the basis \eqref{eq:jm_i} simply after cyclically permuting the labels $x$, $y$ and $z$, without the appearance of any additional phase factors.)

Introduce the notation
\begin{equation}
	\DD^{(j)}_{mn}(h_e)_i = {}_i\bra{jm}\DD^{(j)}(h_e)\ket{jn}_i
	\label{eq:Dg_i}
\end{equation}
for the matrix elements of the normalized Wigner matrix $\DD^{(j)}(h_e) = \sqrt{2j+1}D^{(j)}(h_e)$ in the basis \eqref{eq:jm_i}. Imposing the choice \eqref{eq:n=j}--\eqref{eq:m=j} on the holonomies in the $x$- and $y$-directions as well, with respect to the bases appropriate to these directions, we are led to consider the state
\begin{equation}
	\prod_{i=x, y, z} \DD^{(j_i^+)}_{m_i^+j_i^+}\bigl(h_{e_i^+}\bigr)_i \DD^{(j_i^-)}_{j_i^-n_i^-}\bigl(h_{e_i^-}\bigr)_i
	\label{eq:node-m=j}
\end{equation}
(where $i=z$ refers to the standard basis). Denote this state by $\ket{\Psi_v}$. The calculation which produced \Eq \eqref{eq:M_v-action-z} now shows that
\begin{equation}
	\Bigl(\hat A\bigl(S^a(v)\bigr)^2 - \hat E_a\bigl(S^a(v)\bigr)^2\Bigr)\ket{\Psi_v} = \frac{1}{4}\bigl(j_a^+ + j_a^-\bigr)\ket{\Psi_v}.
	\label{}
\end{equation}
Hence the action of the entire operator \eqref{eq:M_v} on the state \eqref{eq:node-m=j} gives
\begin{equation}
	\hat M_v\ket{\Psi_v} = \frac{1}{4}\sum_a \bigl(j_a^+ + j_a^-\bigr)\widehat{\V_v^{-1}}\ket{\Psi_v}.
	\label{eq:M_v*Psi}
\end{equation}

Now the final step of our calculation is to find the action of the inverse volume operator $\widehat{\V_v^{-1}}$ on the state $\ket{\Psi_v}$. This can be deduced from the results obtained in \cite{Makinen:2020rda}, where the action of the volume operator $\hat V_v$ on a state of the form \eqref{eq:node-m=j} was derived under the assumption that all the spin quantum numbers characterizing the state are large:
\begin{equation}
	j_i^\pm \gg 1. 
	\label{eq:j>>1}
\end{equation}
The derivation was performed in two stages as follows. First, one considers the operator $\hat q_v$ in \Eq \eqref{eq:V_v}. By direct calculation, it can be shown that
\begin{equation}
	\hat q_v\ket{\Psi_v} = \lambda_v\ket{\Psi_v} + {\cal O}(j^2)
	\label{eq:qv*Psi}
\end{equation}
where
\begin{equation}
	\lambda_v = \dfrac{1}{8}\bigl(j_x^+ + j_x^-\bigr)\bigl(j_y^+ + j_y^-\bigr)\bigl(j_z^+ + j_z^-\bigr)
	\label{}
\end{equation}
and the term ${\cal O}(j^2)$ denotes a state which is not of the form \eqref{eq:node-m=j}, but whose norm depends quadratically on the spins. In contrast, the eigenvalue $\lambda_v$ is of order ${\cal O}(j^3)$, so it follows from the assumption \eqref{eq:j>>1} that the second term on the right-hand side of \Eq \eqref{eq:qv*Psi} is small in comparison with the diagonal term. Due to the fact that the action of $\hat q_v$ on the state $\ket{\Psi_v}$ is approximately diagonal, the action of the volume operator $\hat V_v = \sqrt{|\hat q_v|}$ on this state can be deduced by applying standard perturbation theory to establish the relevant part of the spectral decomposition of the operator $\hat q_v$. In this way, as shown in detail in Appendix B of \cite{Makinen:2020rda}, one finds 
\begin{equation}
	\hat V_v\ket{\Psi_v} = \sqrt{\lambda_v}\ket{\Psi_v} + {\cal O}\bigl(\sqrt j\bigr)
	\label{}
\end{equation}
where the norm of the off-diagonal term is again small in comparison with the leading, diagonal term.

In the case presently at hand for us, the arguments presented in Appendix B of \cite{Makinen:2020rda} can equally well be applied to deduce the action of the operator $\widehat{\V_v^{-1}}$ on the state $\ket{\Psi_v}$, given the known action of $\hat q_v$ on this state. The result is
\begin{equation}
	\widehat{\V_v^{-1}}\ket{\Psi_v} = \lambda_v^{-1}\ket{\Psi_v} + {\cal O}\bigl(j^{-5/2}\bigr).
	\label{}
\end{equation}
Going now back to \Eq \eqref{eq:M_v*Psi}, we finally establish that
\begin{equation}
	\hat M_v\ket{\Psi_v} = {\cal O}\bigl(j^{-1/2}\bigl)\ket{\Psi_v} + {\cal O}\bigl(j^{-3/2}\bigr),
	\label{}
\end{equation}
which represents the conclusion of the calculations carried out in this section. We have found a family of states which, while not being exact solutions of the master constraint operator at the node $v$, do satisfy the constraint approximately in the regime of large spin quantum numbers, in the sense that the norm of the state $\hat M_v\ket{\Psi_v}$ approaches zero as the spins are taken larger and larger. This family of states is defined by \Eq \eqref{eq:node-m=j}, and is characterized by the property that the magnetic quantum number of each holonomy at $v$ takes its maximum value with respect to the basis adapted to the direction of the corresponding edge.

The states given by \Eq \eqref{eq:node-m=j} represent a set of approximate solutions of the gauge fixing constraint at the node $v$. To extend the solution to all nodes of a given cubical graph, it suffices to impose the conditions \eqref{eq:n=j}--\eqref{eq:m=j} at every node of the graph. In \Eq \eqref{eq:node-m=j}, this means that both magnetic indices of each holonomy are set to their maximal value. Finally, note that the calculations starting from \Eq \eqref{eq:M_v-action-z} go through in exactly the same way if, instead of \Eqs \eqref{eq:n=j} and \eqref{eq:m=j}, we make the choice $n_z^+ = -j_z^+$ and $m_z^- = -j_z^-$, \ie both magnetic indices now take their minimum value. (However, if one of $n_z^+$ and $m_z^-$ is chosen to be maximal while the other is taken minimal, the off-diagonal terms in \Eq \eqref{eq:M_v-action-z} are not completely eliminated, although the diagonal term behaves in the same way as before.) Hence we will allow both maximal and minimal values of the magnetic quantum numbers, but at any node $v$ of the cubical graph, we require that both magnetic indices corresponding to a given coordinate direction must have the same sign. In this way we arrive at a space of (approximate) solutions of the master constraint operator \eqref{eq:M}, which is summarized in section \ref{sec:H_red} below.

\section{Fundamental structure of quantum-reduced loop gravity}
\label{sec:QRLG}

\subsection{The reduced Hilbert space}
\label{sec:H_red}

We take the Hilbert space of the quantum-reduced model to be the space spanned by the states which were derived in section \ref{sec:solutions} as solutions of the gauge fixing master constraint operator. Thus, a basis on the reduced Hilbert space is given by the states
\begin{equation}
	\prod_{e\in\Gamma} \DD^{(j_e)}_{\tau_ej_e\;\sigma_ej_e}(h_e)_{i_e}.
	\label{eq:basis_reduced}
\end{equation}
The graph $\Gamma$ is a cubical graph, with its edges aligned along the coordinate directions defined by a fixed Cartesian background coordinate system. Note that the calculation of section 3.3 is valid for any cubical graph. Hence one is not necessarily restricted to working with just a single fixed graph; all cubical graphs (corresponding to a specific choice of the background coordinate directions) are included in the Hilbert space of quantum-reduced loop gravity.

All the spin quantum numbers labeling the state \eqref{eq:basis_reduced} are required to be large:
\begin{equation}
	j_e \gg 1
	\label{eq:largej}
\end{equation}
for every edge of the graph $\Gamma$. The labels $\sigma_e$ and $\tau_e$ in the state \eqref{eq:basis_reduced} are sign factors, \ie $\sigma_e = \pm 1$ and $\tau_e = \pm 1$. (The factors $\sigma_e$ and $\tau_e$ belong respectively to the source and target of the edge $e$.) For any given edge, $\sigma_e$ and $\tau_e$ can be chosen independently of each other; however, at each node of the graph the two labels associated with a given coordinate direction must have the same value, as described at the end of section \ref{sec:solutions}. In other words, let $e_i^\pm(v)$ denote an edge which is incident on the node $v$ and aligned in the $x^i$-coordinate direction, with the superscript indicating whether the edge lies before ($-$) or after ($+$) the node $v$ (in the sense defined by the positive direction of the $x^i$-coordinate axis). Then we require that the labels $\sigma_e$ and $\tau_e$ in the state \eqref{eq:basis_reduced} are constrained by the conditions
\begin{equation}
	\sigma_{e_i^+(v)} = \tau_{e_i^-(v)}.
	\label{eq:sigma=tau}
\end{equation}
Regarding the orientation of the graph $\Gamma$, we assume that the orientation is fixed so that every edge of the graph is oriented towards the positive direction of the corresponding background coordinate axis. Note that there is no loss of generality in this assumption, due to the relation
\begin{equation}
	D^{(j)}_{mn}(h_e^{-1}) = (-1)^{m-n}D^{(j)}_{-n\; -m}(h_e)
	\label{eq:D^-1}
\end{equation}
satisfied by the $SU(2)$ representation matrices. This relation shows that a state of the form \eqref{eq:basis_reduced} on a cubical graph of arbitrary orientation is equivalent to a state on a graph with the fixed orientation specified above, only with possibly modified values of some of the sign labels $\sigma_e$ and $\tau_e$.

Apart from the constraint \eqref{eq:sigma=tau}, which has not been considered in the literature of quantum-reduced loop gravity so far, the reduced Hilbert space defined above agrees with the Hilbert space established in the articles originally introducing the quantum-reduced model, together with certain amendments proposed in later work. In particular, the states \eqref{eq:basis_reduced} with opposite signs of the magnetic quantum numbers (\ie $\sigma_e = -\tau_e$) were not included in the Hilbert space considered originally in  \cite{Alesci:2012md, Alesci:2013xd}, but such states were introduced later in \cite{Alesci:2018loi}, where the framework of quantum-reduced loop gravity was applied to the description of spherically symmetric spacetimes. 

\subsection{Action of LQG operators on the reduced Hilbert space}

The relation between the operators of quantum-reduced loop gravity, as originally formulated in the literature of the quantum-reduced model, and the corresponding operators of full loop quantum gravity has been examined in \cite{Makinen:2020rda}. The calculations carried out in \cite{Makinen:2020rda} reveal the following general picture: When an operator $\hat{\cal O}$ on the kinematical Hilbert space of loop quantum gravity is applied on the basis states \eqref{eq:basis_reduced}, the result typically (but not for all possible operators $\hat{\cal O}$) is of the schematic form
\begin{equation}
	\hat{\cal O}\ket{\Psi_0} = f(j)\ket{\Psi_0'} + g(j)\ket{\Phi_\perp}.
	\label{eq:O*psi_red}
\end{equation}
Here $\ket{\Psi_0}$ and $\ket{\Psi_0'}$ are normalized states belonging to the reduced Hilbert space, while $\ket{\Phi_\perp}$ is a normalized state which is orthogonal to the reduced Hilbert space. Moreover, the coefficients $f(j)$ and $g(j)$ are such that, under the assumption of large $j$,
\begin{equation}
	f(j) \gg g(j).
	\label{eq:f>>g}
\end{equation}
Hence the conclusion is that, although the operator $\hat{\cal O}$ does not preserve the reduced Hilbert space exactly, the action of $\hat{\cal O}$ on a reduced state produces a state whose component outside of the reduced Hilbert space is much smaller than the component which lies within the reduced Hilbert space.

In the earlier literature on quantum-reduced loop gravity, the kinematical operators of the model are typically introduced as projections of the corresponding operators of full loop quantum gravity down to the reduced Hilbert space (see \eg \cite{Alesci:2016gub}). Thus, the results obtained in \cite{Makinen:2020rda} provide a justification for this procedure by showing that the terms which get projected out are small in comparison with the terms which are preserved by the projection.

We will now briefly review the results derived in \cite{Makinen:2020rda} pertaining to the holonomy and flux operators, and extend these results to the basis states \eqref{eq:basis_reduced}, in which arbitrary sign factors $\sigma_e$, $\tau_e$ are allowed (in \cite{Makinen:2020rda} only the case $\sigma_e = \tau_e = +1$ was considered). For the flux operator, assuming the surface $S$ intersects the edge $e$ at one of its endpoints, \Eqs \eqref{eq:E*psi} and \eqref{eq:tau_x}--\eqref{eq:tau_z} immediately yield
\begin{equation}
	\hat E_i(S)\DD^{(j)}_{\tau j\;\sigma j}(h_e)_i = \nu(S, e)\times
	\begin{cases}
		\sigma\dfrac{j}{2}\DD^{(j)}_{\tau j\;\sigma j}(h_e)_i & \quad \text{$s(e)$ lies on $S$} \\[2ex]
		\tau\dfrac{j}{2}\DD^{(j)}_{\tau j\;\sigma j}(h_e)_i & \quad \text{$t(e)$ lies on $S$}
	\end{cases}
	\label{eq:E*D_red-diag}
\end{equation}
and
\begin{equation}
	\hat E_k(S)\DD^{(j)}_{\tau j\;\sigma j}(h_e)_i = {\cal O}\bigl(\sqrt j\bigr) \qquad \text{($k\neq i$)}.
	\label{eq:E*D_red-off}
\end{equation}
Hence the dominant component in the action of the flux operator is the one where the internal index matches the direction of the edge on which the operator acts, and the action of this component on the reduced holonomy\footnote{
	We use the term ''reduced holonomy'' to refer to the holonomies which make up the basis state \eqref{eq:basis_reduced}. Thus, the characteristic property of a reduced holonomy associated to an edge $e$ is that both magnetic quantum numbers of the holonomy have the maximal or minimal value with respect to the basis corresponding to the direction of the edge.
}
is diagonal. The remaining components of the flux operator act non-diagonally, producing states which do not belong to the reduced Hilbert space, but these components are suppressed by a factor of $j^{-1/2}$ relative to the diagonal component.

However, note that if the surface intersects the edge at an interior point, the action of the flux operator on a reduced holonomy gives
\begin{equation}
	\hat E_k(S)D^{(j)}_{\tau j\;\sigma j}(h_e)_i = i\nu(S, e)\Bigl(D^{(j)}(h_{e_2})\tau_k^{(j)}D^{(j)}(h_{e_1})\Bigr)_{\tau j\;\sigma j}^{(i)},
	\label{eq:flux-interior}
\end{equation}
where the notation on the right-hand side indicates that the matrix element is taken in the basis $\ket{jm}_i$, \ie $\bigl(M^{(j)}\bigr)_{mn}^{(i)} = {}_i\bra{jm}M^{(j)}\ket{jn}_i$. Here all the matrix elements of the generator $(\tau^{(j)}_i)_{mn}$ enter the result, and the right-hand side of \Eq \eqref{eq:flux-interior} does not reduce to any simple form even under the assumption of large $j$. The flux operator associated to a surface of this kind is therefore an example of an operator for which the result \eqref{eq:O*psi_red} does not hold. In quantum-reduced loop gravity one typically works only with flux operators whose surfaces intersect the cubical graph of the basis states \eqref{eq:basis_reduced} only at the nodes of the graph.

The action of the holonomy operator on a reduced holonomy is given by \Eq \eqref{eq:CGseries} as
\begin{equation}
	\widehat{D^{(s)}_{mn}(h_e)}D^{(j)}_{\tau j\;\sigma j}(h_e) = \sum_k C^{(j\;s\;k)}_{\;\tau j\;m\;m'} C^{(j\;s\;k)}_{\;\sigma j\;n\;n'}D^{(k)}_{m'n'}(h_e),
	\label{eq:D*D_red}
\end{equation}
where
\begin{align}
	m' &= \tau j + m \\
	n' &= \sigma j + n
	\label{}
\end{align}
and we assume that $s = {\cal O}(1)$, while $j\gg 1$ as before. Under these assumptions, it was shown in \cite{Makinen:2020rda} that the Clebsch--Gordan coefficients entering \Eq \eqref{eq:D*D_red} behave as 
\begin{equation}
	C^{(j\;s\;j+m)}_{\;j\;m\;j+m} = 1 + {\cal O}\biggl(\frac{1}{j}\biggr)
	\label{eq:C_large}
\end{equation}
and
\begin{equation}
	C^{(j\;s\;k)}_{\;j\;m\;j+m} = {\cal O}\biggl(\frac{1}{\sqrt j}\biggr) \qquad (k > j+m).
	\label{eq:C_small}
\end{equation}
Using \Eqs \eqref{eq:C_large} and \eqref{eq:C_small} in \Eq \eqref{eq:D*D_red} one finds, considering first the case $\sigma = \tau = +1$,
\begin{align}
	\widehat{D^{(s)}_{mm}(h_e)}\DD^{(j)}_{jj}(h_e) &= \DD^{(j+m)}_{j+m\; j+m}(h_e) + {\cal O}\biggl(\frac{1}{j}\biggr), \label{eq:Dmm_++} \\
	\widehat{D^{(s)}_{mn}(h_e)}\DD^{(j)}_{jj}(h_e) &= {\cal O}\biggl(\frac{1}{\sqrt j}\biggr) \qquad (m\neq n). \label{eq:Dmn_++}
\end{align}
The case $\sigma = \tau = -1$ can be obtained from \Eqs \eqref{eq:Dmm_++} and \eqref{eq:Dmn_++} with the help of the relation \eqref{eq:D^-1}. We have
\begin{align}
	\widehat{D^{(s)}_{mm}(h_e)}\DD^{(j)}_{-j\;-j}(h_e) &= \DD^{(j-m)}_{-j+m\; -j+m}(h_e) + {\cal O}\biggl(\frac{1}{j}\biggr), \label{eq:Dmm_--} \\
	\widehat{D^{(s)}_{mn}(h_e)}\DD^{(j)}_{-j\; -j}(h_e) &= {\cal O}\biggl(\frac{1}{\sqrt j}\biggr) \qquad (m\neq n). \label{eq:Dmn_--}
\end{align}
Thus, the dominant terms in the action of the holonomy operator on a reduced holonomy with $\sigma = \tau$ are given only by the diagonal ($m=n$) components of the operator.

It remains to consider the cases $\sigma = +1, \tau = -1$ and $\sigma = -1, \tau = +1$. These can be deduced from the above results by using the symmetry properties of the Clebsch--Gordan coefficients, namely
\begin{equation}
	C^{(j_1\;j_2\;j)}_{-m_1\;-m_2\;-m} = (-1)^{j_1+j_2-j}C^{(j_1\;j_2\;j)}_{m_1\;m_2\;m}
	\label{}
\end{equation}
(see \eg \cite{Khersonskii:1988krb}). We find
\begin{align}
	\widehat{D^{(s)}_{m\;-m}(h_e)}\DD^{(j)}_{j\;-j}(h_e) &= (-1)^{s-m}\DD^{(j+m)}_{j+m\;-j-m}(h_e) + {\cal O}\biggl(\frac{1}{j}\biggr) \\
	\widehat{D^{(s)}_{m\;-m}(h_e)}\DD^{(j)}_{-j\;j}(h_e) &= (-1)^{s-m}\DD^{(j-m)}_{-j+m\;j-m}(h_e) + {\cal O}\biggl(\frac{1}{j}\biggr)
	\label{}
\end{align}
and
\begin{equation}
	\widehat{D^{(s)}_{mn}(h_e)}\DD^{(j)}_{\pm j\;\mp j}(h_e) = {\cal O}\biggl(\frac{1}{\sqrt j}\biggr) \qquad (m\neq -n).
	\label{}
\end{equation}
Hence, in the case of a reduced holonomy with $\sigma = -\tau$, the leading-order terms in the action of the holonomy operator arise from the ''anti-diagonal'' ($m = -n$) components. In all the cases the leading term is again a reduced holonomy, \ie each magnetic index is either maximal or minimal.

\subsection{Reduced operators}

The results summarized in the previous section suggest that operators of quantum-reduced loop gravity can be interpreted as arising from the action of the corresponding operators of full loop quantum gravity on states in the reduced Hilbert space. (Note that the states \eqref{eq:basis_reduced} are elements of the kinematical Hilbert space of loop quantum gravity.) From any operator $\hat{\cal O}$ whose action on the reduced Hilbert space has the form \eqref{eq:O*psi_red}, we can obtain a reduced operator $\hat{\cal O}^R$ simply by discarding the second term on the right-hand side of \Eq \eqref{eq:O*psi_red}, with \Eq \eqref{eq:f>>g} guaranteeing that the discarded term is small in comparison with the term which is kept. The action of the operator $\hat{\cal O}^R$ on the reduced Hilbert space is given by
\begin{equation}
	\hat{\cal O}^R\ket{\Psi_0} = f(j)\ket{\Psi_0'}.
	\label{}
\end{equation}
From the perspective of the quantum-reduced model, the reduced operator $\hat{\cal O}^R$ is therefore a well-defined operator on the reduced Hilbert space, this space being preserved by the action of the operator. On the other hand, comparing the action of the reduced operator to the original operator $\hat{\cal O}$, we can write
\begin{equation}
	\frac{\bigl|\bigl|\hat{\cal O}\ket{\Psi_0} - \hat{\cal O}^R\ket{\Psi_0}\bigr|\bigr|}{\bigl|\bigl|\hat{\cal O}\ket{\Psi_0}\bigr|\bigr|} \ll 1
	\label{}
\end{equation}
which shows that from the perspective of the full theory, the operator $\hat{\cal O}^R$ provides a good approximation of the action of the full operator $\hat{\cal O}$ on states in the reduced Hilbert space.

The action of the reduced flux operator can be read off from \Eqs \eqref{eq:E*D_red-diag} and \eqref{eq:E*D_red-off} as
\begin{equation}
	\hat E_k^R(S)\DD^{(j)}_{\tau j\;\sigma j}(h_e)_i = \delta_{ik}\nu(S, e) \varphi\frac{j}{2}\DD^{(j)}_{\tau j\;\sigma j}(h_e)_i,
	\label{eq:E_red}
\end{equation}
where $\varphi = \sigma$ or $\varphi = \tau$ according to whether the surface $S$ intersects the edge $e$ at $s(e)$ or $t(e)$. The action of the reduced holonomy operator also follows from the results derived in the previous section. From the various cases considered there, we gather
\begin{equation}
	\widehat{D^{(s)}_{mn}(h_e)_i^R}\,\DD^{(j)}_{\tau j\;\sigma j}(h_e)_i =
	\begin{cases}
		\delta_{mn}\DD^{(j+m)}_{j+m\; j+m}(h_e)_i & (\tau, \sigma) = (++) \\[1ex]
		\delta_{mn}\DD^{(j-m)}_{-j+m\; -j+m}(h_e)_i & (\tau, \sigma) = (--) \\[1ex]
		\delta_{m,-n}(-1)^{s-m}\DD^{(j+m)}_{j+m\; -j-m}(h_e)_i & (\tau, \sigma) = (+-) \\[1ex]
		\delta_{m,-n}(-1)^{s-m}\DD^{(j-m)}_{-j+m\; j-m}(h_e)_i & (\tau, \sigma) = (-+)
	\end{cases}
	\label{}
\end{equation}
It can be seen that in the cases where $\sigma = \tau$, the action of the reduced holonomy operator is essentially the multiplication law of the group $U(1)$, where the role of the $U(1)$ quantum number is played by the magnetic quantum number (and not the spin) carried by the holonomy operator.

Other loop quantum gravity operators can be treated in the same way in order to derive their reduced counterparts. The example of the volume operator was analyzed in \cite{Makinen:2020rda}. It was shown that the action of the reduced volume operator on the basis states \eqref{eq:basis_reduced} is diagonal, and is given by 
\begin{equation}
	\hat V_v^R\biggl(\prod_{e\in\Gamma} \DD^{(j_e)}_{\tau_ej_e\;\sigma_ej_e}(h_e)_{i_e}\biggr) = \lambda_v\biggl(\prod_{e\in\Gamma} \DD^{(j_e)}_{\tau_ej_e\;\sigma_ej_e}(h_e)_{i_e}\biggr)
	\label{eq:V_red}
\end{equation}
where the eigenvalue is
\begin{equation}
	\lambda_v = \sqrt{\frac{1}{8}\bigl(j_x^+ + j_x^-\bigr)\bigl(j_y^+ + j_y^-\bigr)\bigl(j_z^+ + j_z^-\bigr)}
	\label{}
\end{equation}
with $j_i^\pm$ denoting the spins on the six edges incident on $v$, as illustrated in \Fig \ref{fig:node}. Note that the reduced volume operator is not constructed by starting with the reduced flux operators defined by \Eq \eqref{eq:E_red} and using them to form a volume operator. Instead, we evaluate the action of the volume operator of the full theory on the reduced Hilbert space and, extracting the term of leading order in the spins, find that this term is an element of the reduced Hilbert space -- in other words, the action of the operator is of the form \eqref{eq:O*psi_red}. The reduced volume operator of \Eq \eqref{eq:V_red} is then obtained by keeping only the leading term and discarding the terms of lower order in $j$.

To conclude our discussion of reduced operators, let us consider the Hamiltonian constraint. In order to obtain a well-defined Hamiltonian operator for the quantum-reduced model, the operator must be constructed in such a way that the terms of leading order in $j$ produced by the action of the operator on states in the reduced Hilbert space are again elements of the reduced Hilbert space. This condition is necessary (and sufficient) to ensure that the reduced Hilbert space is preserved by the resulting reduced Hamiltonian, which is obtained by truncating the full action of the Hamiltonian at leading order in $j$. (If the space spanned by the states \eqref{eq:basis_reduced} is interpreted as a space of quantum states where the gauge conditions encoded in the master constraint operator \eqref{eq:M} are satisfied, then the above condition can, loosely speaking, be interpreted as the condition for the chosen gauge to be preserved under the quantum dynamics generated by the reduced Hamiltonian.)

A concrete example of a Hamiltonian operator which satisfies the condition formulated above can be obtained as follows. As the classical starting point for the construction, we take the classical Hamiltonian constraint in the form
\begin{equation}
	C(N) = \frac{1}{\beta^2}\int_\Sigma d^3x\,N\biggl(\frac{\epsilon\downup{ij}{k}E^a_iE^b_jF_{ab}^k}{\sqrt{|\det E|}} + (1+\beta^2)\sqrt{|\det E|}\Rt\biggr).
	\label{eq:C}
\end{equation}
Here $\beta$ is the Barbero--Immirzi parameter, $N$ is the lapse function, $F_{ab}^i$ is the curvature of the Ashtekar connection $A_a^i$, and $\Rt$ is the Ricci scalar of the spatial manifold $\Sigma$. For the quantization of the Lorentzian part of the Hamiltonian, represented in \Eq \eqref{eq:C} by the term involving the Ricci scalar, we refer to the work carried out in  \cite{Lewandowski:2021iun, Lewandowski:2022xox}. In this pair of articles, an operator representing the integrated Ricci scalar was first constructed for arbitrary kinematical states of loop quantum gravity based on cubical graphs, and then it was established that the reduced Hilbert space is preserved by the action of this operator at leading order in $j$. To deal with the quantization of the Euclidean part of the Hamiltonian, we follow the construction in \cite{Alesci:2015wla} which, when adapted to the quantization of the Euclidean part of the constraint \eqref{eq:C}, gives rise to the following operator (up to a constant numerical factor, which is omitted):
\begin{equation}
	\hat C_E(N) = \sum_{v\in\Gamma}N(v)\widehat{\V_v^{-1}} \sum_{e \nparallel e'\;\text{at $v$}} \epsilon^{ijk}\Tr\Bigl(\tau_k^{(s)}\widehat{D^{(s)}_{\phantom{m}}(h_{\alpha_{ee'}})}\Bigr)\hat J_i^{(v, e)}\hat J_j^{(v, e')}.
	\label{eq:C_E}
\end{equation}
Here the inner sum runs over pairs of edges whose tangent vectors at $v$ are linearly independent, and the operator $\hat J_i^{(v, e)}$ is defined as
\begin{equation}
	\hat J_i^{(v, e)}D^{(j)}_{mn}(h_e) = \begin{cases}
		-iD^{(j)}_{mm'}(h_e)(\tau_i^{(j)})_{m'n} & v = s(e) \\[1ex]
		i(\tau_i^{(j)})_{mm'}D^{(j)}_{m'n}(h_e) & v = t(e)
	\end{cases}
	\label{}
\end{equation}
However, we depart from \cite{Alesci:2015wla} when it comes to specifying the loop $\alpha_{ee'}$, which arises from the regularization of the curvature $F_{ab}^i$. In our case, the operator \eqref{eq:C_E} is applied on a cubical graph, and we take $\alpha_{ee'}$ to be the minimal closed loop formed by the edges $e$, $e'$ and by two other edges of the cubical graph. Now a straightforward calculation can be performed to show that when the operator \eqref{eq:C_E} acts on the basis states \eqref{eq:basis_reduced}, the term of leading order in $j$ does belong to the reduced Hilbert space. Therefore the construction sketched above provides a well-defined Hamiltonian constraint operator for quantum-reduced loop gravity.

\section{Possible extensions and generalizations}
\label{sec:generalizations}

\subsection{$SU(2)$ gauge invariance}

The basis states \eqref{eq:basis_reduced} are not invariant under the local $SU(2)$ gauge transformations given by \Eq \eqref{eq:g*h}. Hence, when working with a model whose state space is spanned by these states, one essentially takes the point of view that since the gauge freedom represented by the Gauss constraint is completely fixed by the gauge choice \eqref{eq:E_off_diag=0}, there should be no trace of $SU(2)$ gauge invariance left in the quantum theory after the gauge fixing constraint has been imposed. Here we will not try to address the conceptual question of whether the correct way to deal with the Gauss constraint in quantum-reduced loop gravity is simply to ignore it, or whether it should be taken into account in some non-trivial way. Instead, we will take this question as a purely technical one, and propose two possible ways in which the Gauss constraint could be incorporated into the framework considered so far in this article.

Following the extended master constraint proposal of Thiemann \cite{Thiemann:2003zv}, we can include the Gauss constraint\footnote{
	In principle there is no obstruction against treating the spatial diffeomorphism constraint in the same way. An operator representing the master diffeomorphism constraint on a cubical graph can be constructed in a relatively straightforward manner, but the form of this operator is much more complicated than the master Gauss constraint considered in this section, and finding a space of solutions of the constraint in explicit form is a non-trivial task, which we have so far not accomplished.
}
as a part of the master constraint by considering the extended master constraint 
\begin{equation}
	M_{\rm ext} = M + M_{\rm Gauss},
	\label{}
\end{equation}
where $M$ is the master constraint for diagonal gauge from section \ref{sec:master_def}, and the master Gauss constraint is defined as
\begin{equation}
	M_{\rm Gauss} = \int_\Sigma d^3x\,\frac{G_iG_i}{\sqrt q}
	\label{eq:M_Gauss}
\end{equation}
with
\begin{equation}
	G_i = \D_aE^a_i = \partial_a E^a_i + \epsilon\downup{ij}{k}A_a^jE^a_k.
	\label{}
\end{equation}
In order to promote the classical functional \eqref{eq:M_Gauss} into an operator, we use a construction similar (but not identical) to the one introduced in \cite{Lewandowski:2021iun} to quantize covariant derivatives of the densitized triad in terms of finite differences of parallel transported flux operators on a cubical graph.

Consider, as in section \ref{sec:master_def}, a partition of the spatial manifold into cubical cells adapted to a given cubical graph. For any node $v$ of the graph, let $S^a_\pm(v)$ $(a = x, y, z)$ denote the six faces of the cell containing $v$, such that the surface $S^a_\pm(v)$ lies in the coordinate plane $x^a = \const$ and the subscript indicates whether the surface is located before or after the node $v$ in the positive direction of the $x^a$-coordinate axis (as in the notation for the edges in \Fig \ref{fig:node}). Now the covariant derivative $\D_aE^a_i$ at $v$ (no summation over $a$ is understood) can be approximated by the discretized variable
\begin{equation}
	\Delta_a E^a_i(v) = \Et_i\bigl(S^a_+(v), v\bigr) - \Et_i\bigl(S^a_-(v), v\bigr)
	\label{}
\end{equation}
where $\Et(S, v)$ is the parallel transported flux variable discussed in section \ref{sec:LQG-operators}. We assume that the parallel transport from any point $x$ on the surface $S^a_\pm(v)$ to the node $v$ is taken first along a straight line from $x$ to the point at which the surface intersects an edge $e$ of the cubical graph, and then from the intersection point to $v$ along $e$.

We then define
\begin{equation}
	\hat M_{\rm Gauss} = \sum_v \widehat{\V_v^{-1}}\sum_{a, i} \bigl(\widehat{\Delta_a E^a_i(v)}\bigr)^2
	\label{}
\end{equation}
as the operator representing the master Gauss constraint \eqref{eq:M_Gauss}. The action of the operator $\widehat{\Delta_a E^a_i(v)}$ (no sum over $a$) on the basis states \eqref{eq:basis_reduced} can be computed with the help of \Eq \eqref{eq:Et-action}. The result is
\begin{equation}
	\widehat{\Delta_a E^a_i(v)}\biggl(\prod_{e\in\Gamma} \DD^{(j_e)}_{\tau_ej_e\;\sigma_ej_e}(h_e)_{i_e}\biggr) =
	\begin{cases}
		\bigl(j_i^+ - j_i^-\bigr)\biggl(\displaystyle\prod_{e\in\Gamma} \DD^{(j_e)}_{\tau_ej_e\;\sigma_ej_e}(h_e)_{i_e}\biggr) & (i = a) \\[1ex]
		{\cal O}\bigl(\sqrt j\bigr) & (i \neq a)
	\end{cases}
	\label{}
\end{equation}
where the notation of \Fig \ref{fig:node} is again used for the spins at the node $v$. It follows that the constraint equation $\hat M_{\rm Gauss}\ket{\Psi} = 0$ is satisfied exactly by basis states fulfilling the condition $j_i^+ = j_i^-$ at every node. However, since solutions of the extended master constraint operator can again be found only in the approximate sense discussed in section \ref{sec:solutions}, there is room for this condition to be relaxed accordingly. Thus, we consider solutions of the extended master constraint to be the basis states \eqref{eq:basis_reduced} where the spins are constrained by the conditions
\begin{equation}
	\bigl|j_i^+ - j_i^-\bigr| \ll j_i^\pm
	\label{}
\end{equation}
which state that the difference between two consecutive spins along a given ''line'' of edges must be small compared to the spins themselves.

Another possible way of addressing gauge invariance would be to apply $SU(2)$ group averaging (see \eg \cite{Ashtekar:1995zh}) in order to obtain a set of gauge invariant states. Since the gauge conditions do not commute with the Gauss constraint, it should be expected that the outcome of the construction will depend on the order of imposing the gauge-fixing constraint and implementing $SU(2)$ gauge invariance. (Note that this question does not arise in the extended master constraint approach discussed above, where the gauge conditions and the Gauss constraint are both combined into a single constraint operator.) We propose that the natural order of performing these operations is suggested by the fact that the master constraint operator defined by \Eqs \eqref{eq:M} and \eqref{eq:M_v} is not gauge invariant, \ie the gauge invariant Hilbert space is not preserved by the action of this operator. Hence it does not seem clear how this operator could be imposed as a constraint on the gauge invariant Hilbert space in a consistent manner. On the other hand, performing an $SU(2)$ group averaging on the Hilbert space spanned by the states \eqref{eq:basis_reduced} is, from the technical standpoint, a well-defined and unambiguous procedure, acting as a projection onto the $SU(2)$ invariant subspace. For this reason we will only consider the order where the gauge-fixing constraint is solved first, and group averaging is then applied to the resulting basis states \eqref{eq:basis_reduced}.

Let us look at the effect of group averaging on a given node $v$ of a reduced basis state. Here it is convenient to depart temporarily from the convention introduced in section section \ref{sec:H_red} concerning the orientation of the edges, and instead use \Eq \eqref{eq:D^-1} to arrange for all the edges incident on $v$ to have an outgoing orientation. Then the relevant part of the state \eqref{eq:basis_reduced} has the form\footnote{
	The expression \eqref{eq:node-part} is obtained by writing $\DD^{(j)}_{\tau j\; \sigma j}(h_e)_i = {}_i\bra{j,\tau j}\DD^{(j)}(h_e)\ket{j, \sigma j}_i$ and omitting the bra vector ${}_i\bra{j,\tau j}$, which is associated not with the node $v$ but with the target node of the edge $e$. For concreteness, we also assume that all the sign factors in the state \eqref{eq:basis_reduced} are positive at $v$ (before adjusting the orientations): $\sigma_{e_i^+(v)} = \tau_{e_i^-(v)} = +1$.
}
\begin{equation}
	\prod_{i = x, y, z} \DD^{(j_i^+)}\bigl(h_{e_i^+}\bigr)\bket{j_i^+, j_i^+}_i \DD^{(j_i^-)}\bigl(h_{\tilde e_i^-}\bigr)\bket{j_i^-, -j_i^-}_i,
	\label{eq:node-part}
\end{equation}
where $\tilde e_i^-$ stands for the edge $e_i^-$ with a reversed orientation. Applying the group averaging operation to the state \eqref{eq:node-part} now gives
\begin{equation}
	\int dg \prod_{i = x, y, z} \DD^{(j_i^+)}\bigl(h_{e_i^+}g\bigr)\bket{j_i^+, j_i^+}_i \DD^{(j_i^-)}\bigl(h_{\tilde e_i^-}g\bigr)\bket{j_i^-, -j_i^-}_i
	\label{eq:node-averaged}
\end{equation}
where the integral is taken over the $SU(2)$ group manifold. Due to the invariance of the Haar measure, the state \eqref{eq:node-averaged} is manifestly invariant under $SU(2)$ gauge transformations acting at the node $v$.

It is helpful at this point to recall the definition of the coherent states of angular momentum \cite{Radcliffe:1971ayi, Perelomov:1971bd}. The coherent state $\ket{j\vec n}$ is constructed from the state of highest weight $\ket{jj}$ by applying a $SU(2)$ rotation $g(\vec n)$ which rotates the $z$-axis into the direction of the unit vector $\vec n \in \R^3$: $\ket{j\vec n} = D^{(j)}\bigl(g(\vec n)\bigr)\ket{jj}$. A closely related object is the Livine--Speziale coherent intertwiner, which was introduced in \cite{Livine:2007vk} as the $SU(2)$-invariant projection of the tensor product state $\ket{j_1\vec n_1}\cdots\ket{j_N\vec n_N}$:
\begin{equation}
	\ket{j_1\cdots j_N; \vec n_1\cdots\vec n_N} = \int dg\,D^{(j_1)}(g)\ket{j_1\vec n_1}\cdots D^{(j_N)}(g)\ket{j_N\vec n_N}.
	\label{eq:coherent-i}
\end{equation}
Referring now back to \Eq \eqref{eq:node-averaged}, note that the states $\ket{j,\pm j}_i$, as defined in \Eq \eqref{eq:jm_i}, are simply the coherent states $\ket{j\vec n}$ corresponding to the choice $\vec n = \pm\hat e_i$ (with $\hat e_x, \hat e_y, \hat e_z$ being the unit vectors along the background coordinate directions). Hence we see that the group integration in \Eq \eqref{eq:node-averaged} combines with the representation matrices and the state vectors, \ie $D^{(j_i^\pm)}(g)\ket{j_i^\pm, \pm j_i^\pm}_i$, to produce precisely the coherent intertwiner \eqref{eq:coherent-i}. The conclusion emerging from the calculation is therefore as follows: The group averaging of the reduced basis state \eqref{eq:basis_reduced} results in a gauge invariant spin network state of the general form \eqref{eq:spinnetwork}, where each node carries a six-valent Livine--Speziale intertwiner, and the unit vectors $\vec n_1, \dots, \vec n_6$ characterizing the intertwiner are the six face normal vectors of a cube.

Here a crucial question, to which we do not currently have a conclusive answer, has to do with the property summarized in \Eq \eqref{eq:O*psi_red}, whereby the reduced Hilbert space is preserved by the action of the operator $\hat{\cal O}$ at leading order in $j$. It is {\em a priori} not clear whether this property continues to be valid after the group averaging operation has been applied. Assuming that $\hat{\cal O}$ is a gauge invariant operator, and letting $\hat{\mathbbm{P}}_G$ denote the group averaging operator (equivalently, the projection operator onto the gauge invariant Hilbert space), we obtain from \Eq \eqref{eq:O*psi_red}
\begin{equation}
	\hat{\cal O}\hat{\mathbbm{P}}_G\ket{\Psi_0} = f(j)\hat{\mathbbm{P}}_G\ket{\Psi_0'} + g(j)\hat{\mathbbm{P}}_G\ket{\Phi_\perp}.
	\label{eq:O*Ppsi_red}
\end{equation}
However, the state vectors appearing in this equation are no longer normalized (as a projection operator, $\hat{\mathbbm{P}}_G$ does not preserve the norm of the state on which it acts, unless this state is already gauge invariant) and a more careful analysis would be needed in order to establish whether the second term on the right-hand side of \Eq \eqref{eq:O*Ppsi_red} is still much smaller than the first term.

\subsection{A generalized basis of reduced spin network states}
\label{sec:half-edges}

The space spanned by the states \eqref{eq:basis_reduced} is clearly not the most general set of solutions of the gauge-fixing master constraint introduced in section \ref{sec:master_def}, although it is a natural set of solutions of the constraint operator, and up to details related to the sign factors $\sigma_e, \tau_e$, it coincides with the reduced Hilbert space considered in earlier literature of quantum-reduced loop gravity. However, the following considerations suggest that this space may not be general enough with a view towards the possible physical applications of the quantum-reduced model. In the classical theory, the gauge conditions $E^a_i = 0$ for $i\neq a$ generally do not restrict any components of the conjugate variable $A_a^i$ to be vanishing. There exist possible classical configurations in which the densitized triad is diagonal but all components of the Ashtekar connection are non-zero. On the other hand, \Eq \eqref{eq:D*D_red} shows that in the action of the reduced holonomy operator $\widehat{D^{(s)}_{mn}(h_e)^R}$ on the basis \eqref{eq:basis_reduced}, only the diagonal $(m=n)$ and anti-diagonal $(m=-n)$ components of the operator give a non-vanishing result. This does not seem to be consistent with the classical expectation that all components of the connection $A_a^i$ could generally be non-vanishing, since for an arbitrary value of the connection $A_a^i$, all matrix elements of the classical holonomy $D^{(s)}_{mn}\bigl(h_e[A]\bigr)$ will in general be non-zero.

One could ask whether this problem might have to do with the restriction to cubical graphs, whose edges are aligned in the background coordinate directions. Considering that one can find many examples of states in the Hilbert space of full loop quantum gravity which are defined on cubical graphs but on which all components of the holonomy operator act non-trivially, it seems unlikely that the vanishing action of the off-diagonal components on the reduced Hilbert space could be explained entirely by the choice of graph. Nevertheless, it is plausible that this issue could be addressed by looking for solutions of the gauge-fixing master constraint on other types of graphs, on which the action of all components of the holonomy would possibly be non-vanishing. This idea will not be pursued further in the present article, but we leave it as a potential topic for future work. (To the author's best knowledge, there have not been any attempts so far in the literature to extend the framework of quantum-reduced loop gravity to non-cubical graphs.)

Instead, we will propose an extension of the standard reduced Hilbert space described in section \ref{sec:H_red}, in which the system of cubical graphs is not entirely discarded, but is merely modified in a way which allows for all components of the reduced holonomy operator to have a non-trivial action on the extended Hilbert space. In searching for such an extension, we would like to leave the structure present at the nodes of the standard basis states \eqref{eq:basis_reduced} unchanged, since this structure essentially follows from the considerations of section \ref{sec:solutions}, and hence it seems difficult to envision any substantial modification of this structure while ensuring that the resulting states continue to be solutions of the gauge-fixing master constraint. This leaves the possibility of introducing some additional structure at the interior of the edges of the graphs on which the standard basis states are defined.

\begin{figure}[t]
	\centering
	\includegraphics[scale=0.18]{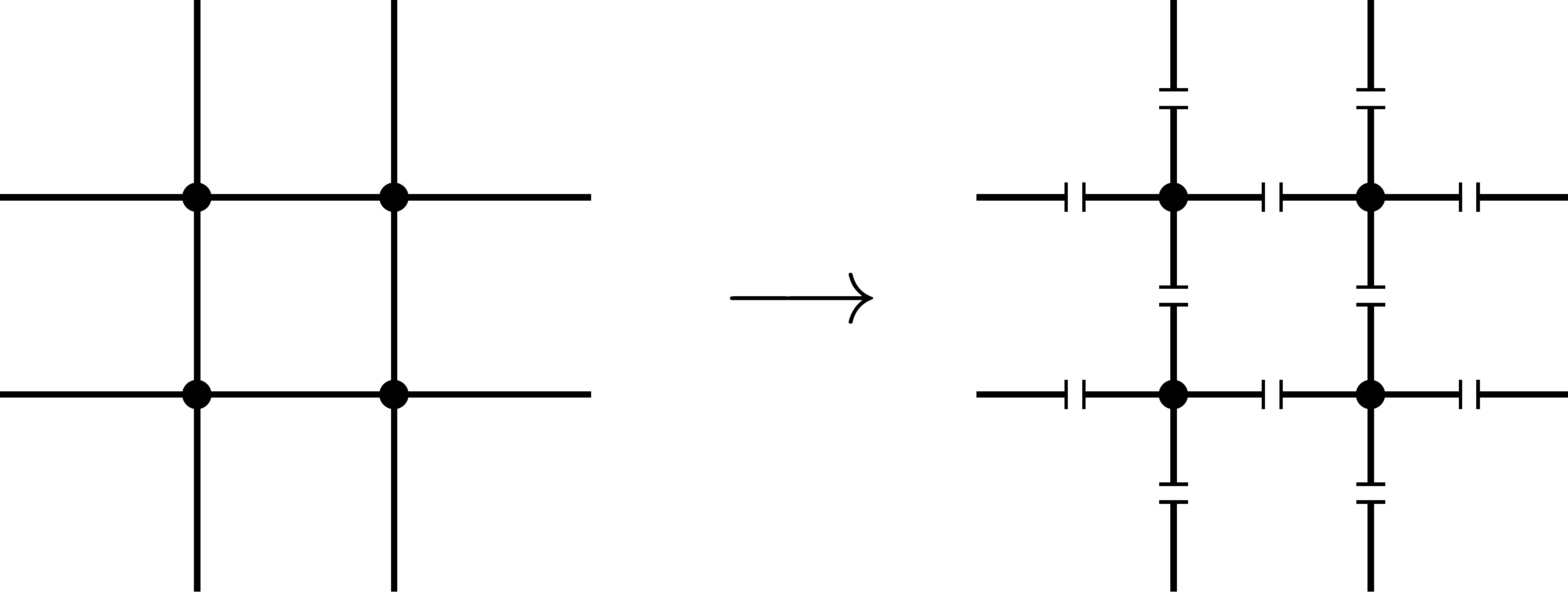}
	\caption{Construction of the generalized basis states. (Third dimension suppressed for clarity.) Each edge of the cubical graph is split into two segments, and the reduced holonomy is replaced with a pair of holonomies, which may carry arbitrary magnetic quantum numbers at the end which is not attached to a proper, non-degenerate node of the graph. Thus, a solid circle in the drawing represents a node where the values of all magnetic indices are maximal or minimal with respect to the appropriate coordinate direction, while a truncated line represents an arbitrary value of the magnetic index.}
	\label{fig:half-edges}
\end{figure}

To this end, consider the basis state \eqref{eq:basis_reduced} and focus on the reduced holonomy corresponding to a given edge $e$. Divide the edge into two segments $e_1$, $e_2$ (so that $e = e_2\circ e_1$) and replace the reduced holonomy $\DD^{(j_e)}_{\tau_ej_e\;\sigma_ej_e}(h_e)_i$ with a product of two holonomies associated with the two segments as follows:
\begin{equation}
	\DD^{(j_e)}_{\tau_e j_e\;\sigma_e j_e}(h_e)_i \; \to \; \DD^{(j_e)}_{\tau_e j_e\;m_e}(h_{e_2})_i\DD^{(j_e')}_{m_e'\;\sigma_e j_e'}(h_{e_1})_i.
	\label{eq:half-edges}
\end{equation}
Here $m_e$ and $m_e'$ are new quantum numbers characterizing the generalized state, and the spins $j_e$ and $j_e'$ may generally be different from each other. From the perspective of standard loop quantum gravity, it may seem somewhat unnatural to assume that the reduced holonomies associated to the two segments can carry different spin quantum numbers. Nevertheless, here it is in fact necessary to allow for this possibility. If the spins on the two segments were required to match, then the Hilbert space spanned by the resulting states would not be closed under the action of the reduced holonomy operator (see \Eq \eqref{eq:D*half-edges} below).

Applying the above procedure to every edge $e$ in the basis states \eqref{eq:basis_reduced}, we arrive at the generalized basis states
\begin{equation}
	\prod_{e\in\Gamma} \DD^{(j_e)}_{\tau_ej_e\;m_e}(h_{e_2})_i \, \DD^{(j_e')}_{m_e'\;\sigma_ej_e'}(h_{e_1})_i.
	\label{eq:basis_generalized}
\end{equation}
In these states, the magnetic quantum numbers associated with every proper six-valent node of the cubical graph $\Gamma$ are $\tau_ej_e$ and $\sigma_ej_e'$, while the quantum numbers $m_e$ and $m_e'$ are located at a ''degenerate'' two-valent node formed by the edges $e_1$ and $e_2$ (see \Fig \ref{fig:half-edges} for a pictorial illustration). When the master constraint operator defined by \Eqs \eqref{eq:M} and \eqref{eq:M_v} is applied on the states \eqref{eq:basis_generalized}, the operator acts non-trivially only on the six-valent nodes; its action on the two-valent nodes vanishes due to the presence of the inverse volume operator $\widehat{{\cal V}_v^{-1}}$. Since the structure at the six-valent nodes of the generalized basis states \eqref{eq:basis_generalized} has not been altered relative to the standard basis states \eqref{eq:basis_reduced}, the generalized states therefore continue to be valid solutions of the gauge-fixing master constraint, as long as the conditions 
\begin{align}
	j_e &\gg 1 \\
	j_e' &\gg 1
	\label{}
\end{align}
are fulfilled for each edge $e$, and the constraint \eqref{eq:sigma=tau} on the sign factors is satisfied at every node.

The action of the holonomy operator on the generalized states \eqref{eq:basis_generalized} can be found by writing the holonomy along the edge $e$ as $h_e = h_{e_2}h_{e_1}$, and then applying \Eq \eqref{eq:CGseries} separately to each of the two segments. Using also \Eqs \eqref{eq:C_large} and \eqref{eq:C_small}, one finds (we display the result only for the case $\sigma_e = \tau_e = +1$; the other cases are not different in any essential way)
\begin{align}
	&\widehat{D^{(s)}_{nn'}(h_e)_i}\DD^{(j)}_{jm}(h_{e_2})_i\DD^{(j')}_{m'j'}(h_{e_1})_i \notag \\[1ex]
	&= \sum_\mu C^{(j\;s\;j+n)}_{m\;\mu\;m+\mu}C^{(j'\;s\;j'+n')}_{m'\;\mu\;m'+\mu}\DD^{(j+n)}_{j+n\;m+\mu}(h_{e_2})_i\DD^{(j'+n')}_{m'+\mu\;j'+n'}(h_{e_1})_i + {\cal O}\bigl(j^{-1/2}\bigr).
	\label{eq:D*half-edges}
\end{align}
The term of leading order in $j$ on the right-hand side of \Eq \eqref{eq:D*half-edges} is a linear combination of states of the form
\begin{equation}
	\DD^{(j)}_{jm}(h_{e_2})_i\DD^{(j')}_{m'j'}(h_{e_1})_i
	\label{}
\end{equation}
so the space of such states is preserved by the action of the reduced holonomy operator obtained from \Eq \eqref{eq:D*half-edges}. Moreover, for any values of $m$ and $n$, the reduced holonomy operator $\widehat{D^{(s)}_{mn}(h_e)_i^R}$ will generally have a non-vanishing action on the basis states \eqref{eq:basis_generalized}.

In the terminology of cylindrical functions, the construction proposed in this section can be seen as a particular kind of refinement of the graph on which the standard basis states \eqref{eq:basis_reduced} are defined. The standard basis states are cylindrical functions on a regular cubical graph containing only six-valent nodes. In contrast, the generalized basis states of \Eq \eqref{eq:basis_generalized} are cylindrical functions on a finer graph, which is obtained by taking a regular cubical graph and introducing a two-valent node at an interior point of every edge. Note that the standard basis states can also be seen as cylindrical functions on this finer graph. (This is achieved by assigning a Kronecker delta at each two-valent node as an intertwiner which contracts the magnetic indices of the holonomies at the node and thus joins the two segments into one whole edge.) Hence the original Hilbert space spanned by the states \eqref{eq:basis_reduced} is a proper subspace of the generalized Hilbert space spanned by the states \eqref{eq:basis_generalized}, so in this sense the generalized Hilbert space is indeed a genuine extension of the standard reduced Hilbert space.

\section{Conclusions}
\label{sec:conclusions}

In this article we constructed and analyzed a constraint operator representing a set of gauge conditions which fix the densitized triad to be diagonal on the kinematical Hilbert space of loop quantum gravity. A gauge fixing to a diagonal triad at the level of the quantum theory is the basic premise of quantum-reduced loop gravity, a physically motivated model of loop quantum gravity proposed by Alesci and Cianfrani in  \cite{Alesci:2012md, Alesci:2013xd}. Hence the work presented in this article can be seen as re-examining the foundations of quantum-reduced loop gravity from an alternative and arguably more systematic standpoint. 

In order to construct a constraint operator encoding the desired gauge conditions, we turned to the master constraint method introduced to loop quantum gravity by Thiemann \cite{Thiemann:2003zv, Thiemann:2005zg} as a technique for the quantization of the Hamiltonian constraint. The gauge conditions aggregated in the resulting master constraint operator do not commute among themselves as quantum operators, despite being compatible with each other in the classical theory. To search for solutions of a constraint of this kind, we followed the approach originally taken by Alesci and Cianfrani. Instead of requiring the constraint equation to be satisfied exactly, one looks for solutions of the constraint among kinematical states characterized by large spin quantum numbers, with the constraint equation being interpreted as a condition to be satisfied approximately in the regime of large spins (the equality becoming exact if one formally takes the limit $j\to\infty$). By deriving a set of solutions of the constraint operator (understood in the above generalized sense) on a cubical graph, we confirmed that the Hilbert space of quantum-reduced loop gravity is recovered in the form considered in earlier literature, up to minor technical details regarding the signs of the magnetic quantum numbers.

Having established a space of solutions of the gauge fixing master constraint, we proceeded to examine the action of the basic operators of loop quantum gravity on this space. We summarized the central results of the earlier article \cite{Makinen:2020rda}, where the relation between the operators of quantum-reduced loop gravity and those of full loop quantum gravity was analyzed, and extended these results to the more general set of states considered in the present article. These calculations show that the Hilbert space of quantum-reduced loop gravity, when viewed as a subspace of the kinematical Hilbert space of full loop quantum gravity, is approximately preserved by the action of many fundamental operators of loop quantum gravity. More precisely, the action of a loop quantum gravity operator on a state in the reduced Hilbert space typically produces a state which lies predominantly in the reduced Hilbert space, in the sense that the component of this state orthogonal to the reduced Hilbert space is much smaller than the component which belongs to the reduced Hilbert space. Operators for quantum-reduced loop gravity can therefore be derived from the corresponding operators of full loop quantum gravity by dropping the small terms which do not belong to the reduced Hilbert space. From the perspective of the quantum-reduced model, the reduced operator obtained in this way is a well-defined operator on the Hilbert space of the model, whereas from the point of view of the full theory the action of the reduced operator can be seen as a good approximation of the action of the full operator on the reduced Hilbert space.

Expressing the gauge conditions for a diagonal triad in terms of a precisely formulated constraint operator provides an opportune starting point for considering various possible generalizations of the construction presented in this article. We showed that the master constraint for diagonal gauge can be straightforwardly extended to include the Gauss constraint, making it possible to implement a notion of $SU(2)$ gauge invariance on the reduced Hilbert space. Another way to address the question of gauge invariance would be to directly apply a $SU(2)$ group averaging map on the reduced basis states. However, for the time being it is not known whether the group averaging operation preserves the decomposition of the action of operators on the reduced Hilbert space into a large term belonging to the reduced Hilbert space and a small term lying outside of this space.

We further proposed a basis of generalized reduced states, making use of the fact that the master constraint operator (in both its non-extended and extended forms) acts only on the nodes of a given graph, so there is room to modify the definition of the standard reduced basis states away from the nodes. The need to look for a generalization of the standard reduced Hilbert space was motivated by the observation that only the diagonal and ''anti-diagonal'' components of the reduced holonomy operator have a non-vanishing action on the standard reduced states. This feature appears to be in conflict with the classical expectation that fixing the off-diagonal components of the densitized triad to zero should not necessarily constrain any components of the Ashtekar connection to vanish. For this reason, it is unclear whether the standard formulation of the reduced Hilbert space can correctly capture all the degrees of freedom contained in the Ashtekar connection in physical situations where off-diagonal components of the connection are relevant. On the space spanned by the generalized reduced states, all components of the reduced holonomy operator have a generically non-vanishing action.

Developing a clearer understanding of the physical interpretation of the generalized reduced states and investigating their possible relation with other ideas discussed in the literature of loop quantum gravity and related fields will make an interesting topic for future work. The splitting of a whole edge of a reduced state into two segments carrying arbitrary magnetic quantum numbers at their ''free'' ends seems to bear at least a superficial resemblance to the discussion of the link entanglement of spin network states in \cite{Bianchi2023}. Moreover, if the notion of splitting an edge in the construction of the generalized reduced states is taken in a very literal sense, and is interpreted to mean that the connectivity information encoded in the graph of the reduced state is also discarded, one would end up with a collection of individual, disconnected six-valent nodes with open-ended edges. This picture is very similar to the so-called condensate states studied in the group field theory approach to quantum gravity (see \eg \cite{Gielen:2016dss, Oriti:2016acw}).

\subsection*{Acknowledgments}

This work was funded by National Science Centre, Poland through grant no. 2022\slash 44\slash C\slash ST2\slash 00023. For the purpose of open access, the author has applied a CC BY 4.0 public copyright license to any author accepted manuscript (AAM) version arising from this submission.

\printbibliography

@article{Alesci:2012md,
    author = "Alesci, Emanuele and Cianfrani, Francesco",
    title = "{A new perspective on cosmology in Loop Quantum Gravity}",
    eprint = "1210.4504",
    archivePrefix = "arXiv",
    primaryClass = "gr-qc",
    doi = "10.1209/0295-5075/104/10001",
    journal = "EPL",
    volume = "104",
    number = "1",
    pages = "10001",
    year = "2013"
}

@article{Alesci:2013xd,
    author = "Alesci, Emanuele and Cianfrani, Francesco",
    title = "{Quantum-Reduced Loop Gravity: Cosmology}",
    eprint = "1301.2245",
    archivePrefix = "arXiv",
    primaryClass = "gr-qc",
    doi = "10.1103/PhysRevD.87.083521",
    journal = "Phys. Rev. D",
    volume = "87",
    number = "8",
    pages = "083521",
    year = "2013"
}

@article{Alesci:2013xya,
    author = "Alesci, Emanuele and Cianfrani, Francesco and Rovelli, Carlo",
    title = "{Quantum-Reduced Loop-Gravity: Relation with the Full Theory}",
    eprint = "1309.6304",
    archivePrefix = "arXiv",
    primaryClass = "gr-qc",
    doi = "10.1103/PhysRevD.88.104001",
    journal = "Phys. Rev. D",
    volume = "88",
    pages = "104001",
    year = "2013"
}

@article{Alesci:2014rra,
    author = "Alesci, Emanuele and Cianfrani, Francesco",
    title = "{Loop quantum cosmology from quantum reduced loop gravity}",
    eprint = "1410.4788",
    archivePrefix = "arXiv",
    primaryClass = "gr-qc",
    doi = "10.1209/0295-5075/111/40002",
    journal = "EPL",
    volume = "111",
    number = "4",
    pages = "40002",
    year = "2015"
}

@article{Alesci:2014uha,
    author = "Alesci, Emanuele and Cianfrani, Francesco",
    title = "{Quantum Reduced Loop Gravity: Semiclassical limit}",
    eprint = "1402.3155",
    archivePrefix = "arXiv",
    primaryClass = "gr-qc",
    doi = "10.1103/PhysRevD.90.024006",
    journal = "Phys. Rev. D",
    volume = "90",
    number = "2",
    pages = "024006",
    year = "2014"
}

@article{Alesci:2015wla,
    author = {Alesci, E. and Assanioussi, M. and Lewandowski, J. and M\"akinen, I.},
    title = "{Hamiltonian operator for loop quantum gravity coupled to a scalar field}",
    eprint = "1504.02068",
    archivePrefix = "arXiv",
    primaryClass = "gr-qc",
    doi = "10.1103/PhysRevD.91.124067",
    journal = "Phys. Rev. D",
    volume = "91",
    number = "12",
    pages = "124067",
    year = "2015"
}

@article{Alesci:2016gub,
    author = "Alesci, Emanuele and Cianfrani, Francesco",
    title = "{Quantum Reduced Loop Gravity and the foundation of Loop Quantum Cosmology}",
    eprint = "1602.05475",
    archivePrefix = "arXiv",
    primaryClass = "gr-qc",
    doi = "10.1142/S0218271816420050",
    journal = "Int. J. Mod. Phys. D",
    volume = "25",
    number = "08",
    pages = "1642005",
    year = "2016"
}

@article{Alesci:2016rmn,
    author = "Alesci, Emanuele and Cianfrani, Francesco",
    title = "{Improved regularization from Quantum Reduced Loop Gravity}",
    eprint = "1604.02375",
    archivePrefix = "arXiv",
    primaryClass = "gr-qc",
    month = "4",
    year = "2016"
}

@article{Alesci:2017kzc,
    author = "Alesci, Emanuele and Botta, Gioele and Stagno, Gabriele V.",
    title = "{Quantum reduced loop gravity effective Hamiltonians from a statistical regularization scheme}",
    eprint = "1709.08675",
    archivePrefix = "arXiv",
    primaryClass = "gr-qc",
    doi = "10.1103/PhysRevD.97.046011",
    journal = "Phys. Rev. D",
    volume = "97",
    number = "4",
    pages = "046011",
    year = "2018"
}

@article{Alesci:2018loi,
    author = "Alesci, Emanuele and Bahrami, Sina and Pranzetti, Daniele",
    title = "{Quantum evolution of black hole initial data sets: Foundations}",
    eprint = "1807.07602",
    archivePrefix = "arXiv",
    primaryClass = "gr-qc",
    doi = "10.1103/PhysRevD.98.046014",
    journal = "Phys. Rev. D",
    volume = "98",
    number = "4",
    pages = "046014",
    year = "2018"
}

@article{Alesci:2018qtm,
    author = "Alesci, Emanuele and Barrau, Aur\'elien and Botta, Gioele and Martineau, Killian and Stagno, Gabriele",
    title = "{Phenomenology of Quantum Reduced Loop Gravity in the isotropic cosmological sector}",
    eprint = "1808.10225",
    archivePrefix = "arXiv",
    primaryClass = "gr-qc",
    doi = "10.1103/PhysRevD.98.106022",
    journal = "Phys. Rev. D",
    volume = "98",
    number = "10",
    pages = "106022",
    year = "2018"
}

@article{Alesci:2019pbs,
    author = "Alesci, Emanuele and Bahrami, Sina and Pranzetti, Daniele",
    title = "{Quantum gravity predictions for black hole interior geometry}",
    eprint = "1904.12412",
    archivePrefix = "arXiv",
    primaryClass = "gr-qc",
    doi = "10.1016/j.physletb.2019.134908",
    journal = "Phys. Lett. B",
    volume = "797",
    pages = "134908",
    year = "2019"
}

@article{Alesci:2019sni,
    author = "Alesci, Emanuele and Botta, Gioele and Luzi, Giovanni and Stagno, Gabriele V.",
    title = "{Bianchi I effective dynamics in Quantum Reduced Loop Gravity}",
    eprint = "1901.07140",
    archivePrefix = "arXiv",
    primaryClass = "gr-qc",
    doi = "10.1103/PhysRevD.99.106009",
    journal = "Phys. Rev. D",
    volume = "99",
    number = "10",
    pages = "106009",
    year = "2019"
}

@article{Alesci:2020zfi,
    author = "Alesci, Emanuele and Bahrami, Sina and Pranzetti, Daniele",
    title = "{Asymptotically de Sitter universe inside a Schwarzschild black hole}",
    eprint = "2007.06664",
    archivePrefix = "arXiv",
    primaryClass = "gr-qc",
    doi = "10.1103/PhysRevD.102.066010",
    journal = "Phys. Rev. D",
    volume = "102",
    number = "6",
    pages = "066010",
    year = "2020"
}

@article{Ashtekar:1993wf,
    author = "Ashtekar, Abhay and Lewandowski, Jerzy",
    title = "{Representation theory of analytic holonomy C* algebras}",
    eprint = "gr-qc/9311010",
    archivePrefix = "arXiv",
    reportNumber = "CGPG-93-8-1",
    month = "11",
    year = "1993"
}

@article{Ashtekar:1994mh,
    author = "Ashtekar, Abhay and Lewandowski, Jerzy",
    title = "{Projective techniques and functional integration for gauge theories}",
    eprint = "gr-qc/9411046",
    archivePrefix = "arXiv",
    reportNumber = "CGPG-94-10-6",
    doi = "10.1063/1.531037",
    journal = "J. Math. Phys.",
    volume = "36",
    pages = "2170--2191",
    year = "1995"
}

@article{Ashtekar:1995zh,
    author = "Ashtekar, Abhay and Lewandowski, Jerzy and Marolf, Donald and Mourao, Jose and Thiemann, Thomas",
    title = "{Quantization of diffeomorphism invariant theories of connections with local degrees of freedom}",
    eprint = "gr-qc/9504018",
    archivePrefix = "arXiv",
    reportNumber = "UCSBTH-95-7",
    doi = "10.1063/1.531252",
    journal = "J. Math. Phys.",
    volume = "36",
    pages = "6456--6493",
    year = "1995"
}

@article{Ashtekar:1997fb,
    author = "Ashtekar, Abhay and Lewandowski, Jerzy",
    title = "{Quantum theory of geometry. 2. Volume operators}",
    eprint = "gr-qc/9711031",
    archivePrefix = "arXiv",
    reportNumber = "CGPG-97-11-1",
    doi = "10.4310/ATMP.1997.v1.n2.a8",
    journal = "Adv. Theor. Math. Phys.",
    volume = "1",
    pages = "388--429",
    year = "1998"
}

@article{Ashtekar:2004eh,
    author = "Ashtekar, Abhay and Lewandowski, Jerzy",
    title = "{Background independent quantum gravity: A Status report}",
    eprint = "gr-qc/0404018",
    archivePrefix = "arXiv",
    doi = "10.1088/0264-9381/21/15/R01",
    journal = "Class. Quant. Grav.",
    volume = "21",
    pages = "R53",
    year = "2004"
}

@book{Ashtekar:2017yom,
    editor = "Ashtekar, Abhay and Pullin, Jorge",
    title = "{Loop Quantum Gravity}: {The First 30 Years}",
    doi = "10.1142/10445",
    isbn = "978-981-320-992-3, 978-981-322-001-0, 978-981-320-993-0",
    publisher = "World Scientific",
    series = "100 Years of General Relativity",
    volume = "4",
    year = "2017"
}

@article{Assanioussi:2020fsz,
    author = "Assanioussi, Mehdi",
    title = "{Graph coherent states for loop quantum gravity}",
    eprint = "2004.08876",
    archivePrefix = "arXiv",
    primaryClass = "gr-qc",
    doi = "10.1103/PhysRevD.101.124022",
    journal = "Phys. Rev. D",
    volume = "101",
    number = "12",
    pages = "124022",
    year = "2020"
}

@inproceedings{Baez:1995md,
    author = "Baez, John C.",
    title = "{Spin networks in nonperturbative quantum gravity}",
    booktitle = "{The Interface of Knots and Physics}",
    eprint = "gr-qc/9504036",
    archivePrefix = "arXiv",
    pages = "167--203",
    month = "4",
    year = "1995"
}

@inbook{Bianchi2023,
	author="Bianchi, Eugenio and Livine, Etera R.",
	editor="Bambi, Cosimo and Modesto, Leonardo and Shapiro, Ilya",
	title="Loop Quantum Gravity and Quantum Information",
	bookTitle="Handbook of Quantum Gravity",
	year="2023",
	publisher="Springer Nature Singapore",
	pages="1--29",
	isbn="978-981-19-3079-9",
	doi="10.1007/978-981-19-3079-9_108-1",
	url="https://doi.org/10.1007/978-981-19-3079-9_108-1"
}

@article{Bianchi:2008es,
    author = "Bianchi, Eugenio",
    title = "{The Length operator in Loop Quantum Gravity}",
    eprint = "0806.4710",
    archivePrefix = "arXiv",
    primaryClass = "gr-qc",
    doi = "10.1016/j.nuclphysb.2008.08.013",
    journal = "Nucl. Phys. B",
    volume = "807",
    pages = "591--624",
    year = "2009"
}

@article{Brunnemann:2007as,
    author = "Brunnemann, Johannes and Rideout, David",
    title = "{Properties of the volume operator in loop quantum gravity. II. Detailed presentation}",
    eprint = "0706.0382",
    archivePrefix = "arXiv",
    primaryClass = "gr-qc",
    reportNumber = "IMPERIAL-TP-2007-DR-02",
    doi = "10.1088/0264-9381/25/6/065002",
    journal = "Class. Quant. Grav.",
    volume = "25",
    pages = "065002",
    year = "2008"
}

@article{Brunnemann:2007ca,
    author = "Brunnemann, Johannes and Rideout, David",
    title = "{Properties of the volume operator in loop quantum gravity. I. Results}",
    eprint = "0706.0469",
    archivePrefix = "arXiv",
    primaryClass = "gr-qc",
    reportNumber = "IMPERIAL-TP-2007-DR-01",
    doi = "10.1088/0264-9381/25/6/065001",
    journal = "Class. Quant. Grav.",
    volume = "25",
    pages = "065001",
    year = "2008"
}

@article{Dupuis:2010jn,
    author = "Dupuis, Maite and Livine, Etera R.",
    title = "{Lifting SU(2) Spin Networks to Projected Spin Networks}",
    eprint = "1008.4093",
    archivePrefix = "arXiv",
    primaryClass = "gr-qc",
    doi = "10.1103/PhysRevD.82.064044",
    journal = "Phys. Rev. D",
    volume = "82",
    pages = "064044",
    year = "2010"
}

@article{Gan:2022mle,
    author = "Gan, Wen-Cong and Ongole, Geeth and Alesci, Emanuele and An, Yang and Shu, Fu-Wen and Wang, Anzhong",
    title = "{Understanding quantum black holes from quantum reduced loop gravity}",
    eprint = "2206.07127",
    archivePrefix = "arXiv",
    primaryClass = "gr-qc",
    doi = "10.1103/PhysRevD.106.126013",
    journal = "Phys. Rev. D",
    volume = "106",
    number = "12",
    pages = "126013",
    year = "2022"
}

@article{Gielen:2016dss,
    author = "Gielen, Steffen and Sindoni, Lorenzo",
    title = "{Quantum Cosmology from Group Field Theory Condensates: a Review}",
    eprint = "1602.08104",
    archivePrefix = "arXiv",
    primaryClass = "gr-qc",
    reportNumber = "IMPERIAL-TP-2016-SG-1",
    doi = "10.3842/SIGMA.2016.082",
    journal = "SIGMA",
    volume = "12",
    pages = "082",
    year = "2016"
}

@article{Giesel:2006uj,
    author = "Giesel, K. and Thiemann, T.",
    title = "{Algebraic Quantum Gravity (AQG). I. Conceptual Setup}",
    eprint = "gr-qc/0607099",
    archivePrefix = "arXiv",
    reportNumber = "AEI-2006-058",
    doi = "10.1088/0264-9381/24/10/003",
    journal = "Class. Quant. Grav.",
    volume = "24",
    pages = "2465--2498",
    year = "2007"
}

@article{Han:2005km,
    author = "Han, Muxin and Huang, Weiming and Ma, Yongge",
    title = "{Fundamental structure of loop quantum gravity}",
    eprint = "gr-qc/0509064",
    archivePrefix = "arXiv",
    doi = "10.1142/S0218271807010894",
    journal = "Int. J. Mod. Phys. D",
    volume = "16",
    pages = "1397--1474",
    year = "2007"
}

@book{Khersonskii:1988krb,
    author = "Khersonskii, V. K. and Moskalev, A. N. and Varshalovich, D. A.",
    title = "{Quantum Theory Of Angular Momentum}",
    doi = "10.1142/0270",
    isbn = "978-981-4415-49-1, 978-9971-5-0107-5",
    publisher = "World Scientific Publishing Company",
    year = "1988"
}

@article{Lewandowski:2021iun,
    author = {Lewandowski, Jerzy and M\"akinen, Ilkka},
    title = "{Scalar curvature operator for models of loop quantum gravity on a cubical graph}",
    eprint = "2110.10667",
    archivePrefix = "arXiv",
    primaryClass = "gr-qc",
    doi = "10.1103/PhysRevD.106.046013",
    journal = "Phys. Rev. D",
    volume = "106",
    number = "4",
    pages = "046013",
    year = "2022"
}

@article{Lewandowski:2022xox,
	author = {Lewandowski, Jerzy and M\"akinen, Ilkka},
	title = "{Scalar curvature operator for quantum-reduced loop gravity}",
	eprint = "2211.04826",
	archivePrefix = "arXiv",
	primaryClass = "gr-qc",
	doi = "10.1103/PhysRevD.107.126017",
	journal = "Phys. Rev. D",
	volume = "107",
	number = "12",
	pages = "126017",
	year = "2023"
}

@article{Livine:2007vk,
    author = "Livine, Etera R. and Speziale, Simone",
    title = "{A New spinfoam vertex for quantum gravity}",
    eprint = "0705.0674",
    archivePrefix = "arXiv",
    primaryClass = "gr-qc",
    reportNumber = "PI-QG-45",
    doi = "10.1103/PhysRevD.76.084028",
    journal = "Phys. Rev. D",
    volume = "76",
    pages = "084028",
    year = "2007"
}

@article{Makinen:2020rda,
    author = {M\"akinen, Ilkka},
    title = "{Operators of quantum-reduced loop gravity from the perspective of full loop quantum gravity}",
    eprint = "2004.00309",
    archivePrefix = "arXiv",
    primaryClass = "gr-qc",
    doi = "10.1103/PhysRevD.102.106010",
    journal = "Phys. Rev. D",
    volume = "102",
    number = "10",
    pages = "106010",
    year = "2020"
}

@article{Olmedo:2018ohq,
    author = "Olmedo, Javier and Alesci, Emanuele",
    title = "{Power spectrum of primordial perturbations for an emergent universe in quantum reduced loop gravity}",
    eprint = "1811.04327",
    archivePrefix = "arXiv",
    primaryClass = "gr-qc",
    doi = "10.1088/1475-7516/2019/04/030",
    journal = "JCAP",
    volume = "04",
    pages = "030",
    year = "2019"
}

@article{Oriti:2016acw,
    author = "Oriti, Daniele",
    title = "{The universe as a quantum gravity condensate}",
    eprint = "1612.09521",
    archivePrefix = "arXiv",
    primaryClass = "gr-qc",
    doi = "10.1016/j.crhy.2017.02.003",
    journal = "Comptes Rendus Physique",
    volume = "18",
    pages = "235--245",
    year = "2017"
}

@article{Perelomov:1971bd,
    author = "Perelomov, A. M.",
    title = "{Coherent states for arbitrary lie groups}",
    doi = "10.1007/BF01645091",
    journal = "Commun. Math. Phys.",
    volume = "26",
    pages = "222--236",
    year = "1972"
}

@article{Radcliffe:1971ayi,
    author = "Radcliffe, J. M.",
    title = "{Some properties of coherent spin states}",
    doi = "10.1088/0305-4470/4/3/009",
    journal = "J. Phys. A",
    volume = "4",
    number = "3",
    pages = "313",
    year = "1971"
}

@article{Rovelli:1994ge,
    author = "Rovelli, Carlo and Smolin, Lee",
    title = "{Discreteness of area and volume in quantum gravity}",
    eprint = "gr-qc/9411005",
    archivePrefix = "arXiv",
    reportNumber = "CGPG-94-11-1",
    doi = "10.1016/0550-3213(95)00150-Q",
    journal = "Nucl. Phys. B",
    volume = "442",
    pages = "593--622",
    year = "1995",
    note = "[Erratum: Nucl.Phys.B 456, 753--754 (1995)]"
}

@article{Rovelli:1995ac,
    author = "Rovelli, Carlo and Smolin, Lee",
    title = "{Spin networks and quantum gravity}",
    eprint = "gr-qc/9505006",
    archivePrefix = "arXiv",
    reportNumber = "CGPG-95-4-4, IASSNS-HEP-95-27",
    doi = "10.1103/PhysRevD.52.5743",
    journal = "Phys. Rev. D",
    volume = "52",
    pages = "5743--5759",
    year = "1995"
}

@book{Rovelli:2004tv,
    author = "Rovelli, Carlo",
    title = "{Quantum gravity}",
    doi = "10.1017/CBO9780511755804",
    publisher = "Cambridge University Press",
    series = "Cambridge Monographs on Mathematical Physics",
    year = "2004"
}

@book{Rovelli:2014ssa,
    author = "Rovelli, Carlo and Vidotto, Francesca",
    title = "{Covariant Loop Quantum Gravity}: {An Elementary Introduction to Quantum Gravity and Spinfoam Theory}",
    isbn = "978-1-107-06962-6, 978-1-316-14729-0",
    publisher = "Cambridge University Press",
    series = "Cambridge Monographs on Mathematical Physics",
    year = "2014"
}

@article{Thiemann:2003zv,
    author = "Thiemann, Thomas",
    title = "{The Phoenix project: Master constraint program for loop quantum gravity}",
    eprint = "gr-qc/0305080",
    archivePrefix = "arXiv",
    reportNumber = "AEI-2003-047, PI-2003-003",
    doi = "10.1088/0264-9381/23/7/002",
    journal = "Class. Quant. Grav.",
    volume = "23",
    pages = "2211--2248",
    year = "2006"
}

@article{Thiemann:2005zg,
    author = "Thiemann, Thomas",
    title = "{Quantum spin dynamics. VIII. The Master constraint}",
    eprint = "gr-qc/0510011",
    archivePrefix = "arXiv",
    reportNumber = "AEI-2005-152",
    doi = "10.1088/0264-9381/23/7/003",
    journal = "Class. Quant. Grav.",
    volume = "23",
    pages = "2249--2266",
    year = "2006"
}

@book{Thiemann:2007pyv,
    author = "Thiemann, Thomas",
    title = "{Modern Canonical Quantum General Relativity}",
    doi = "10.1017/CBO9780511755682",
    isbn = "978-0-511-75568-2, 978-0-521-84263-1",
    publisher = "Cambridge University Press",
    series = "Cambridge Monographs on Mathematical Physics",
    year = "2007"
}

@article{Varadarajan:2021zrk,
    author = "Varadarajan, Madhavan",
    title = "{Euclidean LQG Dynamics: An Electric Shift in Perspective}",
    eprint = "2101.03115",
    archivePrefix = "arXiv",
    primaryClass = "gr-qc",
    doi = "10.1088/1361-6382/abfc2d",
    journal = "Class. Quant. Grav.",
    volume = "38",
    number = "13",
    pages = "135020",
    year = "2021"
}

\end{document}